\documentclass[12pt]{article} 
\begin{document}

\font\dynkfont=cmsy10 scaled\magstep4    \skewchar\dynkfont='60
\def\dynk{\textfont2=\dynkfont}
\def\hr#1,#2;{\dimen0=.4pt\advance\dimen0by-#2pt
              \vrule width#1pt height#2pt depth\dimen0}
\def\vr#1,#2;{\vrule height#1pt depth#2pt}
\def\blb#1#2#3#4#5
            {\hbox{\ifnum#2=0\hskip11.5pt
                   \else\ifnum#2=1\hr13,5.4;\hskip-1.5pt
                   \else\ifnum#2=2\hr13.5,7.6;\hskip-13.5pt
                                  \hr13.5,3.2;\hskip-2pt
                   \else\ifnum#2=3\hr13.7,8.4;\hskip-13.7pt
                                  \hr13,5.4;\hskip-13pt
                                  \hr13.7,2.4;\hskip-2.2pt
                   \else\ifnum#2=7\hr13.7,8.4;\hskip-13.7pt
                                  \hr13.2,6.4;\hskip-13.2pt
                                  \hr13.2,4.4;\hskip-13.2pt
                                  \hr13.7,2.4;\hskip-2.2pt
                   \else\ifnum#2=8\hr13.7,8.4;\hskip-13.7pt
                                  \hr13.2,6.4;\hskip-13.2pt
                                  \hr13.2,4.4;\hskip-13.2pt
                                  \hr13.7,2.4;\hskip-8.2pt$\rangle$\hskip-2.0pt
                   \else\ifnum#2=4\hr13.5,7.6;\hskip-13.5pt
                                  \hr13.5,3.2;\hskip-8pt$\rangle$\hskip-1.8pt
                   \else\ifnum#2=5\hr30,5.4;\hskip-1.5pt         
                   \else\ifnum#2=6\hr13.7,8.4;\hskip-13.7pt
                                  \hr13,5.4;\hskip-13pt
                                  \hr13.7,2.4;\hskip-8.2pt$\rangle$\hskip-2.0pt
                      \fi\fi\fi\fi\fi\fi\fi\fi\fi
                   $#1$
                   \ifnum#4=0
                   \else\ifnum#4=1\hskip-9.2pt\vr22,-9;\hskip8.8pt
                   \else\ifnum#4=2\hskip-10.9pt\vr22,-8.75;\hskip3pt
                                  \vr22,-8.75;\hskip7.1pt
                   \else\ifnum#4=3\hskip-12.6pt\vr22,-8.5;\hskip3pt
                                  \vr22,-9;\hskip3pt
                                  \vr22,-8.5;\hskip5.4pt
                   \else\ifnum#4=5\hskip-9.2pt\vr39,-9;\hskip8.8pt
                                  \fi\fi\fi\fi\fi
                   \ifnum#5=0
                   \else\ifnum#5=1\hskip-9.2pt\vr1,12;\hskip8.8pt
                   \else\ifnum#5=2\hskip-10.9pt\vr1.25,12;\hskip3pt
                                  \vr1.25,12;\hskip7.1pt
                   \else\ifnum#5=3\hskip-12.6pt\vr1.5,12;\hskip3pt
                                  \vr1,12;\hskip3pt
                                  \vr1.5,12;\hskip5.4pt
                   \else\ifnum#5=5\hskip-9.2pt\vr1,29;\hskip8.8pt
                                   \fi\fi\fi\fi\fi

                   \ifnum#3=0\hskip8pt
                   \else\ifnum#3=1\hskip-5pt\hr13,5.4;
                   \else\ifnum#3=2\hskip-5.5pt\hr13.5,7.6;
                                  \hskip-13.5pt\hr13.5,3.2;
                   \else\ifnum#3=3\hskip-5.7pt\hr13.7,8.4;
                                  \hskip-13pt\hr13,5.4;
                                  \hskip-13.7pt\hr13.7,2.4;
                  \else\ifnum#3=7\hskip-5.7pt\hr13.7,8.4;
                                 \hskip-13.2pt\hr13.2,6.4;
                                 \hskip-13.2pt\hr13.2,4.4;
                                  \hskip-13.7pt\hr13.7,2.4;
                  \else\ifnum#3=8\hskip-5.7pt\hr13.7,8.4;
                                 \hskip-13.2pt\hr13.2,6.4;
                                 \hskip-13.2pt\hr13.2,4.4;
                                  \hskip-13.9pt$\langle$\hskip-8pt\hr13.7,2.4;
                   \else\ifnum#3=4\hskip-5.5pt\hr13.5,7.6;
                                  \hskip-13.5pt$\langle$\hskip-8.2pt
                                  \hr13.5,3.2;
                  \else\ifnum#3=5\hskip-5pt\hr30,5.4;
                   \else\ifnum#3=6\hskip-5.7pt\hr13.7,8.4;
                                  \hskip-13pt\hr13,5.4;
                                 \hskip-13.9pt$\langle$\hskip-8.0pt\hr13.7,2.4;
                                 \fi\fi\fi\fi\fi\fi\fi\fi\fi
                   }}
\def\blob#1#2#3#4#5#6#7{\hbox
{$\displaystyle\mathop{\blb#1#2#3#4#5 }_{#6}\sp{#7}$}}
\def\up#1#2{\dimen1=33pt\multiply\dimen1by#1
                  \hbox{\raise\dimen1\rlap{#2}}}
\def\uph#1#2{\dimen1=17.5pt\multiply\dimen1by#1
                  \hbox{\raise\dimen1\rlap{#2}}}
\def\dn#1#2{\dimen1=33pt\multiply\dimen1by#1
                   \hbox{\lower\dimen1\rlap{#2}}}
\def\dnh#1#2{\dimen1=17.5pt\multiply\dimen1by#1
                    \hbox{\lower\dimen1\rlap{#2}}}

\def\rlbl#1{\kern-8pt\raise3pt\hbox{$\scriptstyle #1$}}
\def\llbl#1{\raise3pt\llap{\hbox{$\scriptstyle #1$\kern-8pt}}}
\def\elbl#1{\kern3pt\lower4.5pt\hbox{$\scriptstyle #1$}}
\def\lelbl#1{\rlap{\hbox{\kern-9pt\raise2.5pt\hbox{{$\scriptstyle #1$}}}}}

\def\wht#1#2#3#4{\blob\circ#1#2#3#4{}{}}
\def\blk#1#2#3#4{\blob\bullet#1#2#3#4{}{}}
\def\whtd#1#2#3#4#5{\blob\circ#1#2#3#4{#5}{}}
\def\blkd#1#2#3#4#5{\blob\bullet#1#2#3#4{#5}{}}
\def\whtu#1#2#3#4#5{\blob\circ#1#2#3#4{}{#5}}
\def\drwng#1#2#3{\hbox{$\vcenter{ \offinterlineskip{
  \hbox{\phantom{}\kern6pt{$\circ^{\elbl{#3}}$}}
  \kern-2.5pt\hbox{$\Biggr/$}\kern-0.5pt
  \hbox{\phantom{}\kern-5pt$\circ^{ \elbl{#1}}$}
  \kern-3.0pt\hbox{$\Biggr\backslash$}
  \kern-1.5pt\hbox{\phantom{}\kern6pt{$\circ^{\elbl{#2}}$}}  } }$}}

\def\rde#1#2#3{\raisebox{.5pt}{\hbox{\phantom{}\kern-4pt\hbox{$\vcenter
{\offinterlineskip\hbox{
               \raise 4.5pt\hbox{\vrule height0.4pt width13pt depth0pt}
                \kern-1pt\vbox{ \hbox{\drwng{#1}{#2}{#3}}} }}$  }} }}

\def\andsev#1.#2.#3.#4.#5.{\dynk \whtu0100{#1}\whtu1100{#2}\cdots%
                           \whtu1100{#3}\cdots%
                           \whtu1100{#4}\whtu1000{#5}}
\def\andfour#1.#2.#3.#4.{\dynk \whtu0100{#1}\whtu1100{#2}\cdots%
                           \whtu1100{#3}\whtu1000{#4}}
\def\ddcnst#1.#2.#3.#4.#5{\dynk \whtu0100{#1}\whtu1100{#2}\cdots%
                          \whtu1100{#3}\cdots%
                           \whtu1400{#4}\whtu2000{#5}}

\def\eddanirs#1.#2.#3.#4.#5.{\dynk \whtu0100{#1}\whtu1100{#2}%
                           \cdots\whtu1100{#3}\whtu1200{#4}\whtu4000{#5}}
\def\dddnu#1.#2.#3.#4.#5.#6.{\hbox{$\vcenter{\hbox
         {\dynk\hbox{$ \whtu0100{#1}\whtu1100{#2}\cdots%
          \whtu1100{#3}\rde{#4}{#5}{#6} $}}  }$}}                           
                                                                                %
                                                                                                    %
\renewcommand{\theequation}{\arabic{section}.\arabic{equation}} 

\def\sqr#1#2{{\vcenter{\hrule height.#2pt
      \hbox{\vrule width.#2pt height#1pt \kern#1pt
          \vrule width.#2pt}
      \hrule height.#2pt}}}
\newcommand{\ssquare}{{\mathchoice{\sqr34}{\sqr34}{\sqr{2.1}3}{\sqr{1.5}3}}}
\newcommand{\square}{{\mathchoice{\sqr84}{\sqr84}{\sqr{5.0}3}{\sqr{3.5}3}}}
\newfont{\elevenmib}{cmmib10 scaled\magstep1}
\newfont{\cmssbx}{cmssbx10 scaled\magstep3}                     
\newcommand{\preprint}{                                         
            \begin{flushleft}                                   
            \elevenmib Yukawa\, Institute\, Kyoto               
            \end{flushleft}\vspace{-1.3cm}                      
            \begin{flushright}\normalsize  \sf                  
            YITP-97-28\\ hep-th./9706034 \\ June 1997       
            \end{flushright}}                                   
\newcommand{\Title}[1]{{\baselineskip=26pt \begin{center} 
            \Large   \bf #1 \\ \ \\ \end{center}}}                
\newcommand{\Author}{\begin{center}\large \bf                   
            Hong-Chen Fu\footnote[1]{On leave of absence from   
            Institute of Theoretical Physics, Northeast         
            Normal University, Changchun 130024, P.R.China.     
            }\ \              
            and Ryu Sasaki\end{center}}                        
\newcommand{\Address}{\begin{center} \it                        
            Yukawa Institute for Theoretical Physics, Kyoto     
            University,\\ Kyoto 606-01, Japan \end{center}}     
\newcommand{\Accepted}[1]{\begin{center}{\large \sf #1}\\       
            \vspace{1mm}{\small \sf Accepted for Publication}   
            \end{center}}                                       
\baselineskip=20pt

\preprint
\thispagestyle{empty}										%
 
\bigskip
\bigskip
\bigskip
\Title{ Probability Distributions and Coherent States of \\
 $B_r$, $C_r$ and  $D_r$ Algebras }
\Author

\Address
\vspace{2cm}

\begin{abstract}
\noindent 
A new approach to probability theory based on quantum mechanical and 
Lie algebraic ideas is proposed and developed.
The underlying fact is the observation that the coherent states
of the Heisenberg-Weyl, $su(2)$, $su(r+1)$, $su(1,1)$ and $su(r,1)$ 
algebras in certain symmetric (bosonic) representations
give the ``probability amplitudes'' (or the ``square roots'')
of the well-known Poisson, binomial, multinomial, negative binomial 
and negative multinomial distributions in probability theory.
New probability distributions are derived based on coherent states of
the classical algebras $B_r$, $C_r$ and $D_r$ in symmetric  
representations.
These new probability distributions are simple generalisation of the
multinomial distributions with some added new features reflecting the 
quantum and Lie algebraic construction.
As byproducts, simple  proofs and interpretation    of addition theorems of
Hermite polynomials are obtained from the `coordinate' representation 
of the (negative) multinomial states.
In other words, these addition theorems are higher rank counterparts
of the well-known generating function of Hermite polynomials, which is
essentially the `coordinate' representation of the ordinary 
(Heisenberg-Weyl) coherent state.
\\ \\

\end{abstract}

\newpage

\section{Introduction}
\setcounter{equation}{0}

Quantum theory is one of the greatest achievements in  twentieth 
century physics.
It has changed the fundamental structure of physics, material science 
and also influenced various disciplines, in particular  biological
(genetic) science and philosophy.
Quantum theory dictates that at the microscopic level
 nature is not governed by  causal laws typically exemplified by the
Newtonian equation of motion but by probabilistic laws.
The fundamental ingredient of quantum theory is, however, not the 
probability itself but the probability amplitude which obeys a certain 
equation of motion and the square of which gives appropriate 
probabilities.

In the present paper we report on an attempt to apply {\em quantum 
theory ideas} to  {\em probability theory itself}.
This, we believe, will provide  new perspectives on probability 
theory and hopefully will enrich the long-established and rather 
mature science.
The first step would be to associate certain ``probability amplitudes''
to some typical probability distributions of  classical 
probability theory.
In a broader perspective, this problem belongs to the paradigm of 
``square roots''. The Dirac equation is obtained as a ``square root''
of the Klein-Gordon equation. The creation and annihilation operators 
can be considered as ``square roots'' of the harmonic oscillator 
hamiltonian.
Of course such a ``square root'' can never be unique. It depends on 
the formulation.
It turns out that the `{\em coherent states}' 
\cite{schro,klau,Glau,noch}\
in quantum optics and
the so-called `generalised coherent states'\footnote{In this paper we 
call them simply coherent states.}\cite{Pere1,Pere}
 associated with various Lie 
algebras could be identified as certain ``probability amplitudes''.
For example, the coherent states associated with the Heisenberg-Weyl 
algebra, $su(2)$ \cite{Rad,stol}, $su(r+1)$ \cite{Gil,KSDR}\ and 
$su(1,1)$ \cite{Pere1,Gil,Gerry,fus3,fus6}\ $su(r,1)$ \cite{fus6} algebras in 
totally symmetric (bosonic) representations could well be interpreted 
as ``probability amplitudes'' for the Poisson, binomial, multinomial 
and negative binomial, negative multinomial distributions in
probability theory, respectively \cite{fus6,fus7}.
This also means, in turn, that these typical discrete {\em probability
distributions are characterised in terms of Lie algebras (groups)
and their representations.\/} The relationship between the Poisson 
distribution and the ordinary coherent states is  well-known and 
that of the binomial distribution and the $su(2)$ coherent states 
is also known, but to a lesser degree.
The characterisation of the negative binomial (multinomial) 
distributions by Lie-algebra representations, which is believed to be 
new, has been reported in our previous work \cite{fus6,fus7}.

The second step is to extract useful information (predictions) from the
characterisation ``probability amplitudes\,=\,coherent states''.
One would naturally ask `what would be the probability distributions 
associated with the other Lie algebras and/or other representations?'
In the present paper we mainly address the problems in this step.
We choose the classical Lie algebras, $B_r$, $C_r$ and $D_r$ in Cartan 
notation (or $so(2r+1)$, $sp(2r)$ and $so(2r)$ algebra, respectively)
and construct the coherent states in the totally symmetric (bosonic)
representations. This gives rise to new probability distributions, to 
be denoted as $B_r$ multinomial distributions, etc.
One reason for choosing the symmetric representations is that they are 
supposed to give closest analogs of the classical probability 
distributions, like the multinomial distribution.
Another reason is the relative ease of the calculation and 
presentation.

The third step would be to discuss the time evolution (stochastic 
process) based not on the probability itself but on the ``probability 
amplitude'' in the spirit of quantum theory \cite{books}.
This would be the subject of our future publication.

\bigbreak
This paper is organised as follows.
In section two we explain the basic idea of introducing the 
``probability amplitude'' by taking the simplest and well-known example of
the Poisson distribution and derive the ordinary coherent state. 
This section is meant for wider readership.
In section three we discuss the ``probability amplitudes'' for the
binomial and multinomial distributions, the coherent states of 
$A_1$ ($su(2)$) and $A_r$ ($su(r+1)$) algebras in a slightly 
different way from our previous work \cite{fus7}.
The representation theory aspects of these algebras are emphasised
in order to facilitate the transition to the other algebras treated in 
later sections.
As  new material in this section we discuss the $x$ (coordinate) 
representation of these coherent states. 
Based on new expressions of the $A_1$ and $A_r$ coherent states, which 
have straightforward interpretations of ``probability amplitudes''
for the binomial and multinomial distributions, we obtain a simple
(quantum theoretical) proof and interpretation of addition theorems
of the Hermite polynomials describing the number states of 
harmonic oscillators. This is  analogous to the well-known 
fact that the coordinate representation of the coherent state of the 
Heisenberg-Weyl group gives the generating function of Hermite 
polynomials.
In sections four, five and six, we derive new probability distributions
associated with the totally symmetric (bosonic) representations of
the $C_r$, $B_r$ and $D_r$ algebras, respectively.
These are the first and simplest results of the second step of the
``quantum theory of probability'' mentioned above. Since the Dynkin 
diagram of $C_r$ is obtained from that of $A_{2r-1}$ by folding,
the $C_r$ coherent states  resemble closely  those of the $A_{2r-1}$
algebra.
However, the obtained probability distributions, to be denoted as the 
$C_r$ multinomial distributions, have markedly different features 
from the ordinary multinomial distributions, reflecting the different 
weight space structures of the $C_r$ and $A_{2r-1}$ algebras.
The probability distributions associated with the symmetric 
representations of $B_r$ and $D_r$ algebras have also new and 
interesting features.
Since $B_r$ Dynkin diagram is obtained from that of $D_{r+1}$ by 
folding, these probability distributions are somewhat related.
Section seven is devoted to a summary of results.
In the Appendix we give a simple proof and interpretation of
another type of addition theorems of Hermite polynomials based on the
$x$ representation of $su(1,1)$ and $su(r,1)$ coherent states.
The formula is known as generalised Mehler formula but is not found
in the standard mathematics reference texts.
This time the summation includes infinite number of terms reflecting the 
infinite dimensionality of the irreducible unitary representations of
these non-compact algebras.

\section{``Quantum Theory of Probability'': An Example}
\setcounter{equation}{0}

Let us begin with the naive idea of associating ``probability
amplitude'' to a probability distribution.
In other words, we explain how to give some meaning to a 
``square root'' of a probability distribution by taking the simplest
example of the Poisson distribution.
Throughout this paper we consider only  discrete probability 
distributions $P$ parametrised by a set of integers.
A probability distribution parametrised by one non-negative integer 
$n$ is completely specified by a set of non-negative numbers 
satisfying the conditions of unit total probability:
\begin{equation}
    P_n\ge0, \quad \sum_{n=0}^\infty P_n=1.
    \label{probdis}
\end{equation}
For a quantum theory let us introduce
 a Hilbert space  $\cal H$ with an orthonormal basis
$|n\rangle$, $n=0,1,2,\ldots,$
\begin{equation}
    \langle m|n\rangle=\delta_{m\,n},
    \label{ortho}
\end{equation}
satisfying the completeness relation
\begin{equation}
    I=\sum_{n=0}^\infty|n\rangle\langle n|,
    \label{comp}
\end{equation}
in which $I$ on the left hand side is the identity operator.
Our objective is to find a normalised state
$|\psi\rangle$ in $\cal H$ such that its transition amplitudes
$\langle n|\psi\rangle$  give rise to the probability distribution:
\begin{equation}
    |\langle n|\psi\rangle|^2=P_n,\quad n=0,1,2,\ldots.
    \label{amppro}
\end{equation}
Then by using the completeness relation one obtains
\begin{equation}
    |\psi\rangle=\sum_{n=0}^\infty|n\rangle\langle n|\psi\rangle=
    \sum_{n=0}^\infty e^{i\delta_n}\sqrt{P_n}|n\rangle,
    \label{pnamp}
\end{equation}
in which the phase $\delta_n$ is arbitrary. Thus far the Hilbert 
space is unspecified.

Let us choose as $\cal H$ the Hilbert space of one of the simplest 
quantum systems, the {\em harmonic oscillator}. 
It is described by the
annihilation and creation operators $a$ and $a^\dagger$ satisfying
the commutation relation
\begin{equation}
    [a,a^\dagger]=1.
    \label{aadag}
\end{equation}
(Throughout this paper Planck's constant $\hbar$ is set to unity.)
Then the orthonormal basis is simply given by
\begin{equation}
    |n\rangle={(a^\dagger)^n\over\sqrt{n!}}|0\rangle,
    \quad n=0,1,2,\ldots,
    \label{numsta}
\end{equation}
in which $|0\rangle$ is the vacuum state characterised by the 
condition
\begin{equation}
    a|0\rangle=0.
    \label{vac}
\end{equation}

The well-known Poisson distribution describing random processes
occurring in a time (space) sequence is
\begin{equation}
    P_n(\alpha)=e^{-\alpha^2}{\alpha^{2n}\over{n!}},\quad n=0,1,2,\ldots.
    \label{poi}
\end{equation}
For example, the number of radio-active decay particles emitted from
a sample in a fixed time ($t$) is known to obey this distribution,
$\alpha^2\propto t$. 
Then the quantum state $|\psi(\alpha)\rangle$ (``probability amplitude'') 
corresponding to
the Poisson distribution (\ref{poi}) is easily obtained 
(we set $\delta_n=0$):
\begin{equation}
    |\psi(\alpha)\rangle=e^{-\alpha^2/2}\sum_{n=0}^\infty{\alpha^n\over\sqrt{n!}}
    |n\rangle.
    \label{cohe1}
\end{equation}
If we substitute the definition of the number state in terms of the 
creation operator, we obtain a closed form
\begin{equation}
    |\psi(\alpha)\rangle=e^{-\alpha^2/2}e^{\alpha 
    a^\dagger}|0\rangle=e^{\alpha(a^\dagger-a)}|0\rangle,
    \label{cohe2}
\end{equation}
and the last formula is obtained by using the 
Baker-Campbell-Hausdorff (B-C-H)
formula
\begin{displaymath}
    e^{A+B}=e^Ae^Be^{-{1\over2}[A,B]}
\end{displaymath}
for the case $[A,B]$ commutes with $A$ and $B$.
This state was first introduced by Schr\"odinger \cite{schro} 
and discussed by many 
authors \cite{klau,Glau,noch} under the name `coherent state' 
which was coined by Glauber in quantum optics.
The coherent state has many other characterisations. 
\begin{enumerate}
    \item  It is an eigenstate of the annihilation operator:
    \begin{displaymath}
        a|\psi(\alpha)\rangle=\alpha|\psi(\alpha)\rangle.
    \end{displaymath}

    \item  It is a minimum uncertainty state:
    \begin{displaymath}
        \langle\Delta x^2\rangle\langle\Delta p^2\rangle=1/4.
    \end{displaymath}
    in which $x=(a^\dagger+a)/\sqrt2$, $p=i(a^\dagger-a)/\sqrt2$ are
    the corresponding coordinate and momentum of the oscillator.
    Heisenberg's uncertainty principle dictates that 
\begin{displaymath}
        \langle\Delta x^2\rangle\langle\Delta p^2\rangle\ge1/4,
    \end{displaymath}
    for arbitrary states.
    \item It is obtained by applying a unitary operator (known as the 
    displacement operator)
    \begin{displaymath}
        e^{\alpha(a^\dagger-a)}
    \end{displaymath}
    to the vacuum state. Such unitary operators form a (unitary) 
    representation of the Heisenberg-Weyl group. 
\end{enumerate}
The last characterisation is generalised by many authors and the 
concept of the coherent states associated with various Lie algebras 
(groups) is now well established. Thus starting from a rather naive 
idea of introducing ``probability amplitude'' for the Poisson 
distribution we have arrived at the concept of the coherent states,
a rather solid subject in quantum theory and the representation theory of 
Lie algebras (groups). As we have shown in previous publications 
\cite{fus6,fus7},
the relationship between coherent states and certain probability 
amplitudes is neither coincidental nor superficial but essential.
As we will briefly review in the next section, 
the ``probability amplitudes'' for the well-known binomial and 
multinomial distributions are the coherent states of $su(2)$ and
$su(r+1)$ algebras in the totally symmetric (bosonic) representations.
The same assertion holds for the negative binomial and negative 
multinomial distributions and the corresponding algebras are
$su(1,1)$ and $su(r,1)$, the non-compact counterparts of $su(2)$ and 
$su(r+1)$.

\section{Coherent States of $A_r$ algebra }
\setcounter{equation}{0}
\subsection{Binomial States}

Let us continue along the line of argument of introducing
 ``probability amplitudes'' for classical probability distributions.
Here we consider the binomial distribution:
\begin{equation}
    B_{(n_0,n_1)}(\eta;M)={M \choose n_1}\eta^{2n_1}(1-\eta^2)^{n_0},\quad
    n_0+n_1=M,\quad \eta\in \mathbf{R},
    \label{bindis}
\end{equation}
which describes probability distribution of $M$ Bernoulli trials of
success (probability $\eta^2$) and failure (probability $1-\eta^2$).
Here $n_1$ is the number of successes and $n_0$ failures.
As a Hilbert space let us choose the Fock space generated by two 
independent bosonic oscillators:
\begin{eqnarray}
    [a_j,a_k^\dagger] & = & \delta_{jk},\quad 
[a_j,a_k]=[a^\dagger_j,a^\dagger_k]=0,\quad j,k=0,1,\nonumber\\
    |n_0,n_1\rangle & = & {(a_0^\dagger)^{n_0}(a_1^\dagger)^{n_1}
    \over{\sqrt{n_0!n_1!}}}|0\rangle,\quad a_j|0\rangle=0,\quad j=0,1,
    \label{twobosfock}
\end{eqnarray}
and restrict the total number to $M$ (integer)
\begin{equation}
    n_0+n_1=M.
    \label{binsum}
\end{equation}
Let us denote by $|\eta;M\rangle$ the ``square root'' of the binomial 
distribution within this finite ($M+1$) dimensional  Hilbert space.
Following the same steps as in the previous section, we arrive at
a simple expression:
\begin{eqnarray}
    |\eta;M\rangle & = & \sum_{n_0+n_1=M}|n_0,n_1\rangle\langle n_0,n_1
                      |\eta;M\rangle                   
    \nonumber \\
     & = & 
     \sum_{n_0+n_1=M}{\sqrt{M!}\over{\sqrt{n_0!n_1!}}}\eta^{n_1}
                (1-\eta^2)^{n_0/2}|n_0,n_1\rangle
    \nonumber  \\
     & = & {1\over\sqrt{M!}} \sum_{n_0+n_1=M}{M!\over{n_0!n_1!}}(\eta 
     a_1^\dagger)^{n_1}(\sqrt{1-\eta^2}a_0^\dagger)^{n_0}|0\rangle
    \nonumber  \\
     & = & {1\over\sqrt{M!}}\left(\sqrt{1-\eta^2}a_0^\dagger+\eta 
     a_1^\dagger\right)^M|0\rangle,
    \label{binst1}
\end{eqnarray}
which shows clearly that the ``transition amplitude'' for each
possible result $\langle n_0,n_1|\eta;M\rangle$ is actually obtained 
by the binomial expansion.

\bigbreak
The next step is to identify $|\eta;M\rangle$ as a coherent state. Let 
us recall the  realisation of $su(2)$ algebra in terms of two 
bosonic oscillators:
\begin{eqnarray}
    J_+  =  a_0^\dagger a_1,\quad J_-&=&a_1^\dagger a_0,\quad 
    J_0={1\over2}(a_0^\dagger a_0-a_1^\dagger a_1),
    \nonumber  \\
   \    [J_+,J_-] & = & 2J_0,\quad [J_0,J_\pm]=\pm J_\pm.
    \label{schrel}
\end{eqnarray}
Obviously the restricted two boson Fock space provides the irreducible 
(spin $M/2$) representation of $su(2)$ corresponding to the Young 
diagram
\begin{displaymath}
    \square\square\square \cdots \square\square\ \ M\ {\rm boxes}.
\end{displaymath}
Its normalised highest weight state is
\begin{equation}
    |M,0\rangle={1\over\sqrt{M!}}(a_0^\dagger)^M|0\rangle,\quad 
    J_+|M,0\rangle=0,\quad J_0|M,0\rangle={M\over2}|M,0\rangle.
    \label{su2high}
\end{equation}
Similarly to the coherent states of the Heisenberg-Weyl group in the 
previous section, $su(2)$ coherent states have the form 
\begin{equation}
    U|\psi_0\rangle, \qquad U\in SU(2).
    \label{basest}
\end{equation}
 These coherent states have ``minimal
uncertainty'' if the `base' state $|\psi_0\rangle$ corresponds to a 
dominant weight, i.e., to the highest weight state or its trajectory 
by the Weyl group \cite{delb}.
Thus without loss of generality we choose $|\psi_0\rangle=|M,0\rangle$.
Since $J_+$ annihilates the highest weight state and $J_0$ does not 
change it, the non-trivial action is by $J_-$ only.
So the un-normalised $su(2)$ coherent state is given by
\begin{equation}
    e^{\xi J_-}|M,0\rangle  =    
    {1\over\sqrt{M!}}e^{\xi a_1^\dagger a_0}(a_0^\dagger)^M|0\rangle=                   %
    {1\over\sqrt{M!}}(a_0^\dagger+\xi a_1^\dagger)^M|0\rangle, 
    \qquad  \xi\in  \mathbf{C}.
    \label{binst2}
\end{equation}
Here use is made of the fact that the oscillator algebra 
$[a_0,a_0^\dagger]=1$ is realised by 
$a_0=\partial/\partial a^\dagger_0$ 
and $a_0^\dagger$. At the last equality, the formal Taylor's theorem
    \begin{equation}
        e^{\alpha{d\over{dx}}}f(x)=f(x+\alpha)
        \label{tay}
    \end{equation}
is used. It is easy to get the normalised coherent state
\begin{equation}
    {1\over{M!}}\left(\sqrt{1-|\eta|^2}a_0^\dagger +\eta 
    a_1^\dagger\right)^M|0\rangle,\qquad \eta=\xi/\sqrt{1+|\xi|^2}\in 
    \mathbf{C},
    \label{binst3}
\end{equation}
which has the same form as the binomial state derived above.
(In order to get complex $\eta$ we only have to choose the phase of 
$\sqrt{B_{(n_0,n_1)}(\eta;M)}$ appropriately.)
Thus we have shown that the ``probability amplitude'' of the
binomial distribution is the $su(2)$ coherent state.

\subsection{Multinomial States}
In this subsection we discuss the relationship between the multinomial
distributions and the $A_r$ coherent states \cite{FKSF}, which has been 
demonstrated in some detail in our previous paper \cite{fus7}.
Here we give a simpler and clearer proof of the correspondence with 
more emphasis on the Lie algebraic structures (i.e., roots and 
weights) which would be useful for comparison with the results of the 
other algebras discussed in later sections.

\bigbreak
The multinomial distribution is
\begin{equation}
    M_\mathbf{n}(\hbox{\boldmath $\eta$}\,;M)={M!\over{n_0!\cdots 
    n_r!}}\ \eta_0^{2n_0}\eta_1^{2n_1}\cdots\eta_r^{2n_r},
    \quad n_0+n_1+\cdots+n_r=M,
    \label{mulnomdis}
\end{equation}
in which
\begin{equation}
    \mathbf{n}=(n_0,n_1,\ldots,n_r),\quad 
    \eta_0^2=1- \hbox{\boldmath $\eta$}^2,\quad
    0< \hbox{\boldmath $\eta$}^2=\eta_1^2+\cdots+\eta_r^2<1,\quad 
    \eta_j\in \mathbf{R}, \quad j=0,\ldots,r.
    \label{mulnot}
\end{equation}
As a Hilbert space let us   choose the Fock space generated by
$r+1$ independent bosonic oscillators
\begin{eqnarray}
    [a_j,a_k^\dagger]&=&\delta_{jk},\qquad \qquad a_j| 0\rangle=0,
    \qquad \qquad j=0,1,\ldots,r, 
    \nonumber\\ 
    |\mathbf{n}\rangle&=&{({\mathbf a}^\dagger)^{\mathbf n}\over
    \sqrt{{\mathbf n}!}}| 
    0\rangle,\quad 
    ({\mathbf a}^\dagger)^{\mathbf 
    n}=(a_0^\dagger)^{n_0}(a_1^\dagger)^{n_1}\cdots(a_r^\dagger)^{n_r},
    \quad {\mathbf n}!=n_0!n_1!\cdots n_r!,
    \label{mulosci}
\end{eqnarray}
and restrict the total number to be $M$
\begin{equation}
    n_0+n_1+\cdots+n_r=M.
    \label{rmpart}
\end{equation}
It has the  dimension
\begin{equation}
    {M+r \choose M}={M+r \choose r}.
    \label{ardim}
\end{equation}
Let us denote by $|\hbox{\boldmath $\eta$};M\rangle$ the ``square root''
of the multinomial distribution within this Hilbert space.
Then we obtain in a similar way to the binomial state
\begin{eqnarray}
    |\hbox{\boldmath $\eta$};M\rangle & = & \sum_{n_0+\cdots+n_r=M}
    |n_0,\cdots,n_r\rangle\langle n_0,\cdots,n_r|\hbox{\boldmath $\eta$};M\rangle
    \nonumber  \\
     & = & \sum {\sqrt{M!}\over\sqrt{n_0!\cdots n_r!}}\,\eta_0^{n_0}\cdots
     \eta_r^{n_r}|n_0,n_1,\cdots,n_r\rangle
    \nonumber  \\
     & = & {1\over\sqrt{M!}}\sum{M!\over{n_0!n_1!\cdots 
     n_r!}}\,(\eta_0a_0^\dagger)^{n_0}\cdots(\eta_ra_r^\dagger)^{n_r}
     |0\rangle
    \nonumber  \\
     & = & 
     {1\over\sqrt{M!}}\left(\eta_0a_0^\dagger+\eta_1a_1^\dagger+
     \cdots+\eta_ra_r^\dagger\right)^M|0\rangle.
    \label{multnom}
\end{eqnarray}

\bigbreak
Now let us consider $A_r$ algebra and its representations. Its Dynkin 
diagram is a simple line connecting $r$ vertices. The number 
attached to each vertex corresponds to the name of the simple roots
given below.
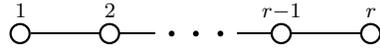
\begin{displaymath}
	\andfour1.2.{r-1}.r.
\end{displaymath}
The simple roots are most conveniently expressed in terms of $r+1$ 
orthonormal vectors in $\mathbf{R}^{r+1}$, $e_j\cdot e_k=\delta_{jk}$,
$j,k=0,1,\ldots,r$:
\begin{equation}
    \alpha_1=e_0-e_1,\quad \alpha_2=e_1-e_2,\quad \cdots,\quad 
    \alpha_r=e_{r-1}-e_r.
    \label{arsimroot}
\end{equation}
Then any root, positive or negative, can be expressed as
\begin{equation}
    e_j-e_k,\qquad j\ne k,
    \label{arroots}
\end{equation}
which is positive if $j<k$ and negative for $j>k$. All the roots have 
the same length.
The fundamental weight vectors, $\{\lambda_j; j=1,\ldots,r\}$, the 
dual basis of the simple root system
\begin{equation}
    2\lambda_j\cdot\alpha_k/\alpha_k^2=\delta_{jk},
    \label{weidef}
\end{equation}
can also be expressed by $\{e_j\}$. For example
\begin{equation}
    \lambda_1={1\over{r+1}}\left(r\,\alpha_1+(r-1)\alpha_2+\cdots+
    \alpha_r\right)=e_0-\left(e_0+e_1+\cdots+e_r\right)/(r+1).
    \label{lam1ar}
\end{equation}
We consider the irreducible representation of $A_r$ with the highest 
weight
\begin{equation}
    \mu=M\lambda_1=M\,e_0-M\left(e_0+e_1+\cdots+e_r\right)/(r+1),
    \label{arhighwei}
\end{equation}
corresponding to the Young diagram
\begin{displaymath}
    \square\square\square \cdots \square\square\ \ M\ {\rm boxes},
\end{displaymath}
which has the same dimension ${M+r\choose r}$ as the restricted
multiboson Fock space introduced above.
Thus this completely symmetric representation can be realised in terms 
of $r+1$ bosonic oscillators.
The weights and the occupation numbers are related one to one, namely the
state $|n_0,n_1,\ldots,n_r\rangle$ has the weight
\begin{equation}
    \mu=\sum_{j=0}^rn_je_j-M\left(e_0+e_1+\cdots+e_r\right)/(r+1).
    \label{arweight}
\end{equation}
All the weight spaces are non-degenerate, i.e., one-dimensional.

If we denote the $A_r$ generators corresponding to the root $e_j-e_k$ by
$X_{(j,-k)}$, we have
\begin{equation}
    X_{(j,-k)}=a_j^\dagger a_k
    \label{argen}
\end{equation}
and
\begin{eqnarray}
    \ [X_{(j,-k)},X_{(k,-l)}]   & = & [a_j^\dagger a_k,a_k^\dagger 
    a_l]=a_j^\dagger a_l=X_{(j,-l)},
    \nonumber  \\
    \ [X_{(j,-k)},X_{(k,-j)}]  & = & H_{(j,k)} \equiv a_j^\dagger 
    a_j-a_k^\dagger a_k.
    \label{arcomrel}
\end{eqnarray}
Here $H_{(j,k)}$ belongs to the Cartan subalgebra. The quadratic 
Casimir operator is 
\begin{equation}
    \mathbf{C}_2={r\over{r+1}}N_{tot}(N_{tot}+r+1),\qquad
    N_{tot}=\sum_{j=0}^r a_j^\dagger a_j,
    \label{arcas}
\end{equation}
which takes the value $rM(M+r+1)/(r+1)$ in the present representation.
The state having the highest weight (\ref{arhighwei}) is
\begin{equation}
    |M,0,\ldots,0\rangle={(a_0^\dagger)^M\over\sqrt{M!}}|0\rangle,
    \label{arhiwest}
\end{equation}
which is annihilated by the generators
\begin{equation}
    X_{(j,k)},\qquad H_{(j,k)}, \qquad j,k=1,\ldots,r,
    \label{armsub}
\end{equation}
forming an $A_{r-1}$ subalgebra. The action of the Cartan subalgebra 
generators $H_{(0,j)}$ does not change the state, either:
\begin{displaymath}
    H_{(0,j)}|M,0,\ldots,0\rangle=M|M,0,\ldots,0\rangle.
\end{displaymath}
Thus the coherent states based on the highest weight state 
(\ref{arhighwei}) are characterised by
\begin{equation}
    SU(r+1)/U(1)\times SU(r)=\mathbf{CP}^r.
    \label{cpr}
\end{equation}
Among the generators belonging to $\mathbf{CP}^r$, only those
\begin{equation}
    X_{(j,-0)}=a_j^\dagger a_0,\qquad j=1,\ldots,r
    \label{aract}
\end{equation}
have non-trivial action on the highest weight state (\ref{arhighwei}).
Thus we find, as in the case of the binomial state (\ref{binst2}),
that the un-normalised $A_r$ coherent state is expressed as
\begin{eqnarray}
     &  & e^{\sum_{j=1}^r\xi_jX_{(j,-0)}}|M,0,\ldots,0\rangle
    \nonumber  \\
     & = & {1\over\sqrt{M!}}\,e^{(\sum_{j=1}^r\xi_ja_j^\dagger)a_0}
     (a_0^\dagger)^M|0\rangle
    \nonumber  \\
     & = & {1\over\sqrt{M!}}\left(a_0^\dagger+
     \sum_{j=1}^r\xi_ja_j^\dagger\right)^M|0\rangle, \qquad
     \qquad  \hbox{\boldmath $\xi$}=(\xi_1,\ldots,\xi_r)\in 
     \mathbf{CP}^r,
    \label{arunmulti}
\end{eqnarray}
in which use has been made of the Taylor expansion theorem (\ref{tay}) 
with $a_0={\partial/\partial a_0^\dagger}$.

\bigbreak
The normalised $A_r$ coherent state in the totally symmetric 
representation is given by
\begin{equation}
    |\hbox{\boldmath $\eta$};M\rangle={1\over\sqrt{M!}}\left(\eta_0a_0^\dagger+
         \sum_{j=1}^r\eta_ja_j^\dagger\right)^M|0\rangle,
         \qquad \eta_j=\xi_j/\sqrt{1+|\hbox{\boldmath $\xi$}|^2}\in {\mathbf 
         C},\quad
         \eta_0=\sqrt{1-|\hbox{\boldmath $\eta$}|^2},
    \label{arnormmult}
\end{equation}
which has the same form as the multinomial state 
$|\hbox{\boldmath $\eta$};M\rangle$ derived above.
As in the binomial state case the ``transition amplitude''
$\langle n_0,\ldots,n_r|\hbox{\boldmath $\eta$};M\rangle$ to each
number state (or weight state 
$\langle \mu_1,\ldots,\mu_r|\hbox{\boldmath $\eta$};M\rangle$) is
simply obtained by multinomial expansion.

\subsection{Coordinate Representation and Addition Theorems of
Hermite Polynomials I}

In this subsection we consider the `coordinate representation' of
the multinomial state (\ref{arnormmult}). This representation is 
useful in quantum optics. It also gives a simple proof and 
interpretation of the following addition theorem of Hermite
polynomials (see, for example, \cite{Feriet} and p196 of \cite{emot}):
\begin{eqnarray}
     &  & {(\eta_0^2+\cdots+\eta_r^2)^{M/2}\over{M!}}H_M\left(
(\eta_0x_0+\cdots+\eta_rx_r)/\sqrt{\eta_0^2+\cdots+\eta_r^2}\right)
    \nonumber  \\
     & = & \sum_{n_0+\cdots+n_r=M}{\eta_0^{n_0}\over{n_0!}}\cdots
     {\eta_r^{n_r}\over{n_r!}}H_{n_0}(x_0)\cdots H_{n_r}(x_r).
    \label{addthe1}
\end{eqnarray}
Here $\eta_0$,\ldots,$\eta_r$ are arbitrary complex numbers. 
It should be noted that the left hand side contains 
$\sqrt{\eta_0^2+\cdots+\eta_r^2}$ in even powers only, since  
Hermite polynomials have a definite parity:
\begin{displaymath}
    H_M(-x)=(-1)^MH_M(x). 
\end{displaymath}

\bigbreak
Let us begin with a single boson oscillator
\begin{displaymath}
    [a,a^\dagger]=1.
\end{displaymath}
The coordinate representation of the number state $|n\rangle$ is
\begin{equation}
    \langle x|n\rangle={1\over\sqrt{n!}}\langle x|(a^\dagger)^n|0\rangle=
    {1\over{\pi^{1/4}2^{n/2}\sqrt{n!}}}H_n(x)\,e^{-{1\over2}x^2},
    \label{xrep}
\end{equation}
in which  Hermite polynomial $H_n$ is given by Rodrigues formula:
\begin{equation}
	H_n(x)=(-1)^ne^{x^2}D^ne^{-x^2},\quad D={d\over{dx}}.
	\label{rodr}
\end{equation}
It is well-known that
the generating function of the Hermite polynomials
\begin{equation}
    \sum_{n=0}^\infty{t^n\over{n!}}H_n(x)=e^{-t^2+2tx}
    \label{hergen}
\end{equation}
is essentially the same
as the coordinate representation of the coherent state of the Heisenberg-Weyl 
group (\ref{cohe1}):
\begin{equation}
    \langle x|\psi(\alpha)\rangle=
    e^{-{1\over2}(x-\sqrt2\alpha)^2}/\pi^{1/4},\qquad \alpha\in
    \mathbf{R}.
    \label{cohxrep}
\end{equation}
The coordinate representation of the multinomial state (\ref{arnormmult}) is
simply obtained by expansion ($\eta_1$, \dots,$\eta_r$ are in general 
complex):
\begin{eqnarray}
     &  & \langle x_0,x_1,\ldots,x_r|\hbox{\boldmath $\eta$};M\rangle
    \nonumber \\
     & = & {1\over\sqrt{M!}} \langle x_0,x_1,\ldots,x_r|
     \left(\eta_0a_0^\dagger+\cdots
         +\eta_ra_r^\dagger\right)^M|0\rangle
    \nonumber  \\
     & = & \sqrt{M!}\ {e^{-{1\over2}(x_0^2+\cdots+x_r^2)}\over
     {\pi^{(r+1)/4}2^{M/2}}}\sum_{n_0+\cdots+n_r=M}{\eta_0^{n_0}\over{n_0!}}\cdots
     {\eta_r^{n_r}\over{n_r!}}H_{n_0}(x_0)\cdots H_{n_r}(x_r).
    \label{multexp1}
\end{eqnarray}

Next we consider operators $A$ and $\widetilde{A}$ defined by
\begin{equation}
    A={\eta_0a_0+\cdots+\eta_ra_r\over\sqrt{\eta_0^2+\cdots+\eta_r^2}},
    \qquad
    \widetilde{A}={\eta_0a_0^\dagger
    +\cdots+\eta_ra_r^\dagger\over\sqrt{\eta_0^2+\cdots+\eta_r^2}}.
    \label{AAdag}
\end{equation}
They are not hermitian conjugate of each other but they satisfy the 
same  relations as those of the single oscillator:
\begin{displaymath}
    [A,\widetilde{A}]=1,\qquad A|0\rangle=0,
\end{displaymath}
which are essential for deriving  Hermite polynomials. 
Thus we obtain
\begin{eqnarray}
     &  &\langle x_0,x_1,\ldots,x_r|\hbox{\boldmath $\eta$};M\rangle
    \nonumber  \\
     & = & {(\eta_0^2+\cdots+\eta_r^2)^{M/2}\over{\sqrt{M!}}}
     \langle x_0,x_1,\ldots,x_r|\widetilde{A}^M|0\rangle
    \nonumber  \\
     & = & {(\eta_0^2+\cdots+\eta_r^2)^{M/2}\over{\sqrt{M!}}}
     {e^{-{1\over2}(x_0^2+\cdots+x_r^2)}\over{\pi^{(r+1)/4}2^{M/2}}}H_M\left(
(\eta_0x_0+\cdots+\eta_rx_r)/\sqrt{\eta_0^2+\cdots+\eta_r^2}\right).
    \label{multexp2}
\end{eqnarray}
Comparing (\ref{multexp1}) and (\ref{multexp2}) we obtain the above 
mentioned addition theorem (\ref{addthe1}) of Hermite polynomials,
which is nothing but the multinomial expansion of the multinomial 
state.
In the Appendix we give a proof and interpretation of another type of 
addition theorems of Hermite polynomials based on negative multinomial 
states, i.e., the coherent states of $su(r,1)$ algebra in discrete 
symmetric representations.

\section{$C_r$ Multinomial States}
\setcounter{equation}{0}

Let us proceed to the second step in the study of ``quantum probability''.
In the previous sections we have shown that some of the typical 
discrete probability distributions are characterised by Lie algebras 
through coherent states. Now we reverse the logic and try to
derive new probability distributions starting from Lie algebras and
their representations.
For this we have, in principle, an infinite choice of Lie algebras
and their representations.
Probably most of such new probability distributions are too exotic to
have any practical use at the moment.
However, the great role played by the Poisson, the binomial, the 
multinomial distributions and their ``negative'' (non-compact) 
counterparts makes us expect that the probability distributions
related with the totally symmetric representations of the other 
classical algebras, $B_r$, $C_r$ and $D_r$ could be useful, though
possibly to a lesser degree.
Apart from the Poisson distribution which has only one parameter,
the (negative) multinomial distribution has many parameters,
\hbox{\boldmath $\eta$} and $M$, to give suitable description to 
various statistical phenomena.
The same property is shared by all the probability distributions derived
from the totally symmetric representations of $B_r$, $C_r$ and $D_r$
algebras. We propose to call these coherent states the 
 $B_r$, $C_r$ and $D_r$ {\em multinomial states\/} and the corresponding 
probability distributions the  $B_r$, $C_r$ and $D_r$ {\em multinomial
distributions\/}. 
We start with the $C_r$ case and proceed to $D_r$ and $B_r$ cases,
in the order of increasing complexity.

\subsection{Coherent States}
The Dynkin diagram of $C_r$ is obtained from that of $A_{2r-1}$ by 
folding.

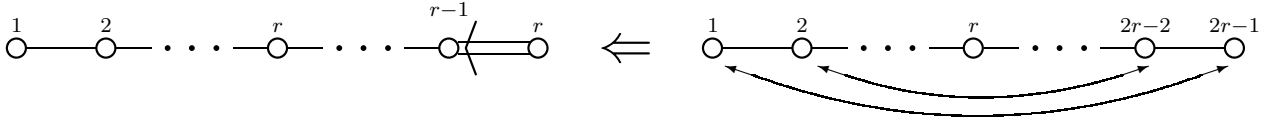
\begin{picture}(450,80)
\put(-30,40){$\ddcnst1.2.r.{r-1}.r\Leftarrow
\andsev1.2.r.{2r-2}.{2r-1}.$}
\put(435,35){\vector(3,1){10}}
\put(265,35){\vector(-3,1){10}}
\qbezier(265,35)(350,4)(435,35)
\put(405,35){\vector(3,1){10}}
\put(300,35){\vector(-3,1){10}}
\qbezier(300,35)(350,18)(405,35)
\end{picture}\\
Its simple roots can be expressed most conveniently in terms of an
orthonormal basis of $\mathbf{R}^r$, $e_j\cdot e_k=\delta_{jk}$,
$j,k=0,\ldots,r$:
\begin{equation}
    \alpha_1=e_1-e_2,\quad \alpha_2=e_2-e_3,\ \cdots,\ 
    \alpha_{r-1}=e_{r-1}-e_r,\quad \alpha_r=2e_r.
    \label{crsimroot}
\end{equation}
The positive roots are
\begin{equation}
    e_j-e_k,\qquad (j<k), \qquad e_j+e_k, \qquad 2e_j.
    \label{crposroot}
\end{equation}
There are $2r(r-1)$ short roots and $2r$ long  roots ($\pm 2e_j$) and the 
dimensions of $C_r$ algebra is $2r^2+r$. The fundamental weights are
\begin{equation}
    \lambda_1=e_1,\qquad \lambda_2=e_1+e_2,\qquad \ldots
    \label{crwei}
\end{equation}
We consider the irreducible representation with the highest weight
\begin{equation}
    \mu=M\lambda_1=Me_1.
    \label{crhighwei}
\end{equation}
Its dimensionality is
\begin{displaymath}
    {M+2r-1 \choose 2r-1}={M+2r-1 \choose M}.
\end{displaymath}
It is the same as the dimension of the restricted 
multiboson ($M$ particle) Fock space
of $A_{2r-1}$ with $2r$ bosonic oscillators:
\begin{equation}
    [a_j,a_k^\dagger]=[b_j,b_k^\dagger]=\delta_{jk},\qquad 
    j,k=1,\ldots,r
    \label{crosci}
\end{equation}
with the number states
\begin{equation}
    |n_1,\ldots,n_r;\bar{n}_1,\ldots,\bar{n}_r\rangle,
    \qquad n_1+\cdots+n_r+\bar{n}_1+\cdots+\bar{n}_r=M,
    \label{crnumst}
\end{equation}
in which $n_j$ ($\bar{n}_j$) is the number of $a_j$ ($b_j$) quanta.

\bigbreak
Similarly to the $A_r$ case, we introduce the following notation for
the generators corresponding to the roots:
\begin{eqnarray}
    X_{(j,-k)} & \Leftrightarrow & e_j-e_k,
    \nonumber  \\
    X_{(j,k)} &  \Leftrightarrow & e_j+e_k, \quad
    X_{(-j,-k)}   \Leftrightarrow  -e_j-e_k,
    \nonumber\\
    X_{(j,j)} &  \Leftrightarrow & 2e_j, \qquad \quad
    X_{(-j,-j)}   \Leftrightarrow  -2e_j.
    \label{crgennot}
\end{eqnarray}
Their forms are
\begin{eqnarray}
    X_{(j,-k)} & = & a_j^\dagger a_k-b_k^\dagger b_j,
    \nonumber  \\
    X_{(j,k)} & = & a_j^\dagger b_k+a_k^\dagger b_j,\quad
    X_{(-j,-k)}  =  b_j^\dagger a_k+b_k^\dagger a_j,
    \nonumber  \\
    X_{(j,j)} & = & a_j^\dagger b_j,\qquad \qquad
    X_{(-j,-j)}  =  b_j^\dagger a_j.
    \label{crgenform}
\end{eqnarray}
It is elementary to check the commutation relations,
for example:
\begin{eqnarray}
    [X_{(j,-k)},X_{(k,-l)}]&=&
    [a_j^\dagger a_k-b_k^\dagger b_j,a_k^\dagger a_l-b_l^\dagger b_k]=
    a_j^\dagger a_l-b_l^\dagger b_j=X_{(j,-l)},
    \nonumber\\
\   [X_{(j,-k)},X_{(k,-j)}]&=&
    a_j^\dagger a_j-b_j^\dagger b_j-a_k^\dagger a_k + b_k^\dagger b_k
    \equiv H_j-H_k,\quad etc.
    \label{crcom1}
\end{eqnarray}
The quadratic Casimir operator is
\begin{equation}
    C_2=N_{tot}(N_{tot}+2r),\qquad 
    N_{tot}=\sum_{j=1}^r(a_j^\dagger a_j+b_j^\dagger b_j),
    \label{crc2}
\end{equation}
which gives $M(M+2r)$ in the present representation.
It is easy to see that each number state belongs to some weight
\begin{equation}
    |n_1,\ldots,n_r;\bar{n}_1,\ldots,\bar{n}_r\rangle \Rightarrow
    \mu=\sum_{j=1}^r(n_j-\bar{n}_j)e_j.
    \label{crnumwei}
\end{equation}
In contradistinction with the $A_{2r-1}$ case this correspondence is 
not 1 to 1.
Some weight spaces are degenerate. For example for $M=4$ and $r=2$,
\begin{displaymath}
    |1,1;1,1\rangle, \quad |2,0;2,0\rangle, \quad |0,2;0,2\rangle
\end{displaymath}
belong to the null weight $\mu=0$.

\bigbreak
As in the case of the binomial states (\ref{basest}) 
we adopt as the `base' state $|\psi_0\rangle$ the highest weight state
\begin{equation}
    |M,0,\ldots,0;0,\ldots,0\rangle=
    {(a_1^\dagger)^M\over\sqrt{M!}}|0\rangle,
    \label{crhiweist}
\end{equation}
which guarantees  ``minimum uncertainty''.
Together with all the positive root generators, it is also 
annihilated by the following generators:
\begin{equation}
    X_{(j,-k)},\quad X_{(j,k)},\quad X_{(-j,-k)},\quad X_{(j,j)},\quad
    X_{(-j,-j)},\quad H_j,\qquad 2\leq j,k\leq r,
    \label{crsub}
\end{equation}
which form a $C_{r-1}$ subalgebra. Likewise the action of the Cartan 
subalgebra generator $H_1$ does not change the highest weight state.
Therefore the $C_r$ multinomial states are parametrised by
\begin{displaymath}
    Sp(2r)/U(1)\times Sp(2(r-1))=\mathbf{CP}^{2r-1},
\end{displaymath}
which also indicates the connection to the $A_{2r-1}$ case.
In fact the generators having non-trivial action on the highest
weight state are
\begin{equation}
    X_{(-1,j)},\quad 2\leq j\leq r\quad \mbox{and} \quad
    X_{(-1,-j)},\quad 1\leq j\leq r.
    \label{crnontri}
\end{equation}
The generators in the first (second) group commute among themselves.
In particular, $X_{(-1,-1)}$ which belongs to the lowest root,
commutes with all the generators in the list (\ref{crnontri}).
The non-commuting pairs among the above generators are
\begin{equation}
    [X_{(-1,j)},X_{(-1,-j)}]=-2X_{(-1,-1)}, \qquad 2\leq j\leq r,
    \label{crnoncom}
\end{equation}
and the resulting generator commutes with all the other generators in 
the list (\ref{crnontri}), as shown above.

\bigbreak
In terms of $2r-1$ complex parameters
\begin{equation}
    \xi_j,\ 2\leq j\leq r, \quad \xi_{-j}, \ 1\leq j\leq r,
    \quad  \hbox{\boldmath $\xi$}=
    (\xi_2,\ldots,\xi_r;\xi_{-1},\ldots,\xi_{-r})\in 
    \mathbf{CP}^{2r-1},
    \label{crpara}
\end{equation}
the un-normalised coherent state is expressed as
\begin{equation}
    e^{C+D}(a_1^\dagger)^M|0\rangle,\quad
    C=\sum_{j=2}^r\xi_jX_{(-1,j)},\quad
    D=\sum_{j=1}^r\xi_{-j}X_{(-1,-j)},
    \label{crcoh1}
\end{equation}
with $[C,D]=2(\sum_{j=2}^r\xi_j\xi_{-j})X_{(-1,-1)}$ commuting with 
$C$ and $D$.
With the help of the B-C-H formula
\begin{displaymath}
    e^{C+D}=e^{C-{1\over2}[C,D]}e^D
\end{displaymath}
and the formal Taylor expansion theorem (\ref{tay}) we arrive at
the following expression of the un-normalised $C_r$ multinomial state
\begin{equation}
    \left(a_1^\dagger+\sum_{j=2}^r\xi_ja_j^\dagger+
    \sum_{j=1}^r\xi_{-j}b_j^\dagger\right)^M|0\rangle,
    \label{crcoh2}
\end{equation}
in which the effects of non-commutativity cancel out exactly.
Therefore the normalised $C_r$ multinomial state is
\begin{equation}
    |\hbox{\boldmath $\eta$};M;C_r\rangle=
    {1\over\sqrt{M!}}\left(\sum_{j=1}^r\eta_ja_j^\dagger
    +\sum_{j=1}^r\eta_{-j}b_j^\dagger\right)^M|0\rangle,
    \label{crcoh3}
\end{equation}
in which
\begin{equation}
    \eta_1=\left(1+\sum_{j=2}^r|\xi_j|^2+\sum_{j=1}^r|\xi_{-j}|^2\right)^{-{1\over2}},
    \quad
    \eta_j=\xi_j\eta_1,\quad \eta_{-j}=\xi_{-j}\eta_1,\qquad 2\leq j\leq 
    r,
    \label{crcohpar}
\end{equation}
satisfying the condition
\begin{displaymath}
    \sum_{j=1}^r(|\eta_j|^2+|\eta_{-j}|^2)=1.
\end{displaymath}
This has exactly the same form as the $A_{2r-1}$ multinomial state.

\subsection{Probability Distribution}
Now we derive the probability distribution from the coherent state,
which has exactly the same form as the $A_r$ multinomial state. So it
predicts the multinomial distribution for the {\em numbers\/} 
$n_1$,\ldots,$\bar{n}_r$
with the  corresponding probabilities $|\eta_1|^2$,\ldots,$|\eta_{-r}|^2$: 
\begin{equation}
    |\langle n_1,\ldots,n_r;\bar{n}_1,\ldots,\bar{n}_r|
    \hbox{\boldmath $\eta$};M;C_r\rangle|^2=
    {M!\over{n_1!\cdots n_r!\bar{n}_1!\cdots 
    \bar{n}_r!}} |\eta_1|^{2n_1}\cdots|\eta_r|^{2n_r}
    |\eta_{-1}|^{2\bar{n}_1}\cdots|\eta_{-r}|^{2\bar{n}_r}.
    \label{crmulpr}
\end{equation}
As remarked above, the $C_r$ states are labeled by the weight
\begin{displaymath}
    \hbox{\boldmath $\mu$}=(\mu_1,\ldots,\mu_r)
\end{displaymath}
which takes positive, zero and negative integer values.
Each weight space has one or many number states which are orthogonal
to each other.
Therefore the $C_r$ multinomial distribution is obtained by summing the
contributions from these number states:
\begin{equation}
    C_{\hbox{\boldmath $\mu$}}(\hbox{\boldmath{$\eta$}};M)=
    \sum_{n_j-\bar{n}_j=\mu_j}
    {M!\over{n_1!\cdots n_r!\bar{n}_1!\cdots 
        \bar{n}_r!}} |\eta_1|^{2n_1}\cdots|\eta_r|^{2n_r}
        |\eta_{-1}|^{2\bar{n}_1}\cdots|\eta_{-r}|^{2\bar{n}_r}.
    \label{crmulprform}
\end{equation}

Let us interpret it in terms of ``picking up balls from a pot''.
The pot contains an infinite number of balls of $r$-different colours.
There are two types of balls for each colour, the 
``{\em positive\/}'' one and ``{\em negative\/}'' one.
Let the probabilities of picking one $j$-th colour ball be 
$\eta_j^2$ for the ``{\em positive\/}'' and $\eta_{-j}^2$ for the 
``{\em negative\/}''.
We pick up total of $M$ balls and ask the probability distribution 
for the ``net'' number of balls (or the ``weight'') for each colour:
$\mu_j=n_j-\bar{n}_j$, $j=1,\ldots,r$.
It is given by the $C_r$ multinomial distribution.
We see that the folding of the $A_{2r-1}$ Dynkin diagram leading to that of 
$C_r$ is very suggestive of this situation.

\section{$D_r$ Multinomial States}
\setcounter{equation}{0}

Here we will derive probability distributions associated with
the symmetric representations of $D_r$ algebra.
They have some new features not present in the multinomial
distributions associated with $A_{2r-1}$ or $C_r$ algebras. 
The Dynkin diagram of $D_r$ algebra with the names of simple roots
attached to the vertices is
shown below. 
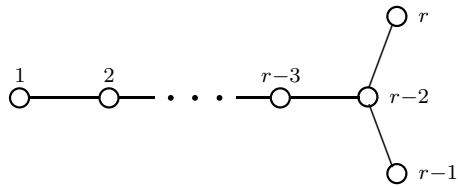
\begin{displaymath}
	\dddnu{1}.{2}.{r-3}.{r-2}.{r-1}.{r}.
\end{displaymath}
The corresponding simple roots are
\begin{equation}
\alpha_1=e_1-e_2,\
\alpha_2=e_2-e_3,\ldots,\alpha_{r-2}=e_{r-2}-e_{r-1},\
\alpha_{r-1}=e_{r-1}-e_r,\ \alpha_r=e_{r-1}+e_r. 
\label{drsimproot}
\end{equation}
The positive roots are all of the same length:
\begin{equation}
e_j-e_k \quad (j<k),\qquad e_j+e_k. 
\label{drposroot}
\end{equation}
The dimension of $D_r$ algebra is $2r^2-r$. The fundamental weights
are
\begin{equation}
\lambda_1=e_1, \quad \lambda_2=e_1+e_2,\ldots,
\label{drsimweu}
\end{equation}
and we consider, as before, the irreducible representation with
highest weight
\begin{equation}
\mu=M\lambda_1=Me_1.
\label{drhiwei}
\end{equation}
Let us denote this representation by $\rho_D^M$ and the corresponding
vector space by $V_D^M$. We know from Weyl's dimension formula
\begin{equation}
\mbox{dim}(V_D^M)={M+2r-3 \choose 2r-3}\times{M+r-1\over{r-1}}.
\label{drdim}
\end{equation}
Let us realise this representation in terms of $2r$ bosons
\begin{displaymath}
a_1,\ldots,a_r,\quad b_1,\ldots,b_r,
\end{displaymath}
and in its restricted Fock space denoted by $F_{2r}^M$, 
\begin{equation}
F_{2r}^M;\quad |n_1,\ldots,n_r;\bar{n}_1,\ldots,\bar{n}_r\rangle,
\qquad n_1+\cdots+n_r+\bar{n}_1+\cdots+\bar{n}_r=M.
\label{f2rfock}
\end{equation}
We have
\begin{equation}
\mbox{dim}(F_{2r}^M)={M+2r-1 \choose 2r-1}={M+2r-1 \choose M}.
\label{f2rdim}
\end{equation}
Comparing (\ref{drdim}) and (\ref{f2rdim}), we find
\begin{eqnarray}
\mbox{dim}(F_{2r}^M)&=&\mbox{dim}(V_D^M)+\mbox{dim}(F_{2r}^{M-2})
\nonumber\\
&=&  \mbox{dim}(V_D^M)+\mbox{dim}(V_D^{M-2})+\cdots,
\label{drdimrel}
\end{eqnarray}
which means that the bosonic Fock space $F_{2r}^M$ contains several
irreducible representations $\rho_D^L$ with different $L$'s.

Let us introduce, as in  the $C_r$ case,  the following
notation for the generators corresponding to the roots:
\begin{eqnarray}
    X_{(j,-k)} & \Leftrightarrow & e_j-e_k,
    \nonumber  \\
    X_{(j,k)} &  \Leftrightarrow & e_j+e_k, \quad
    X_{(-j,-k)}   \Leftrightarrow  -e_j-e_k.
    \label{drgennot}
\end{eqnarray}
Their forms are
\begin{eqnarray}
    X_{(j,-k)} & = & a_j^\dagger a_k-b_k^\dagger b_j,
    \nonumber  \\
    X_{(j,k)} & = & a_j^\dagger b_k-a_k^\dagger b_j,\quad
    X_{(-j,-k)}  =  b_k^\dagger a_j-b_j^\dagger a_k.
    \label{drgenform}
\end{eqnarray}
It is elementary to check the commutation relations,
for example they are (\ref{crcom1}) and:
\begin{eqnarray}
    [X_{(j,-k)},X_{(k,l)}]&=&
    [a_j^\dagger a_k-b_k^\dagger b_j,a_k^\dagger b_l-a_l^\dagger b_k]=
    a_j^\dagger b_l-a_l^\dagger b_j=X_{(j,l)},
    \nonumber\\
\   [X_{(j,k)},X_{(-j,-k)}]&=&
    a_j^\dagger a_j-b_j^\dagger b_j+a_k^\dagger a_k - b_k^\dagger b_k
    \equiv H_j+H_k,\quad etc.
    \label{drcom1}
\end{eqnarray}
The quadratic Casimir operator is
\begin{equation}
    C_2=N_{tot}\left(N_{tot}+2(r-1)\right)-4K^\dagger K,\qquad 
    N_{tot}=\sum_{j=1}^r(a_j^\dagger a_j+b_j^\dagger b_j),
    \label{drc2}
\end{equation}
in which $K$ and $K^\dagger$ are quadratic operators in the
oscillators
\begin{equation}
K=\sum_{j=1}^r a_jb_j,\qquad
K^\dagger=\sum_{j=1}^M a_j^\dagger b_j^\dagger.  
\label{drKdefs}
\end{equation}
They commute with all the above generators including those belonging
to the Cartan subalgebra:
\begin{equation}
[K,X_{\pm(j,\pm k)}]=[K,H_j]=[K^\dagger,X_{\pm(j,\pm
k)}]=[K^\dagger,H_j]=0.
\label{drKcom}
\end{equation} 
In terms of $K^\dagger$ we can express the decomposition of
the bosonic Fock space succinctly:
\begin{equation}
F_{2r}^M=V_D^M\oplus V_D^{M-2}\oplus\cdots V_D^1(V_D^0),
\label{drrepdec}
\end{equation}
in which the vector space $V_D^M$ is obtained from the highest weight
state
\begin{equation}
|M,0,\ldots,0;0,\ldots,0\rangle={(a_1^\dagger)^M\over\sqrt{M!}}|0
\rangle,
\label{drhiwest}
\end{equation}
by applying the negative weight generators successively.
The $j$-th vector space in the right hand side $V_D^{M-2(j-1)}$
is obtained from the highest weight state
\begin{equation} 
{(a_1^\dagger)^{M-2(j-1)}\over\sqrt{(M-2(j-1))!}}\,(K^\dagger)^{j-1}|0
\rangle,
\label{drsubsp}
\end{equation}
by applying the  negative weight generators successively.
It is easy to see that $K$ annihilates all the states in $V_D^M$
\begin{displaymath}
Kv=0,\quad \forall v\in V_D^M,
\end{displaymath}
and we get $C_2=M(M+2(r-1))$ in the highest weight representation
(\ref{drhiwei}),(\ref{drhiwest}).
 It is easy to see that each number state
belongs to some weight
\begin{equation}
    |n_1,\ldots,n_r;\bar{n}_1,\ldots,\bar{n}_r\rangle \Rightarrow
    \mu=\sum_{j=1}^r(n_j-\bar{n}_j)e_j.
    \label{drnumwei}
\end{equation}

\bigbreak
The highest weight state (\ref{drsubsp})
is  annihilated by the following generators belonging to a $D_{r-1}$ subalgebra
\begin{equation}
    X_{(j,-k)},\quad X_{(j,k)},\quad X_{(-j,-k)},\quad
     H_j,\qquad 2\leq j,k\leq r,
    \label{drsub}
\end{equation}
as well as by  all the positive root generators. 
The Cartan 
subalgebra generator $H_1$ does not change the highest weight state.
In other words, the generators having non-trivial action on the highest
weight state are
\begin{equation}
    X_{(-1,j)},\quad  
    X_{(-1,-j)},\quad 2\leq j\leq r.
    \label{drnontri}
\end{equation}
If we denote the compact group corresponding to $D_r$ by $SO(2r)$,
the $D_r$ multinomial states are parametrised by
\begin{displaymath}
    SO(2r)/U(1)\times SO(2(r-1)),
\end{displaymath}
having the dimension
\begin{displaymath}
	4(r-1).
\end{displaymath}
In terms of $2(r-1)$ complex parameters
\begin{equation}
    \xi_j,\quad   \xi_{-j}, \quad 2\leq j\leq r,
    \label{drpara}
\end{equation}
we define a linear combination of the non-trivial generators 
(\ref{drnontri}) as
\begin{equation}
	T=\sum_{j=2}^r \xi_jX_{(-1,j)} +\sum_{j=2}^r\xi_{-j}X_{(-1,-j)}.
	\label{drnontricom}
\end{equation}
It should be noted that all the generators in (\ref{drnontricom}) or
 (\ref{drnontri}) commute among themselves, since the sum of the 
 corresponding roots are not roots any more.
Thus we arrive at the expression of the un-normalised coherent state:
\begin{equation}
    \exp{[T]}(a_1^\dagger)^M|0\rangle=
    \prod_{j=2}^r\exp(\xi_jX_{(-1,j)})
    \prod_{j=2}^r\exp(\xi_{-j}X_{(-1,-j)})(a_1^\dagger)^M|0\rangle.
    \label{drcoh1}
\end{equation}
By repeated use of the formal Taylor expansion theorem (\ref{tay}) we 
obtain
the following explicit form
\begin{equation}
    \left(a_1^\dagger+\sum_{j=2}^r\xi_ja_j^\dagger+
    \sum_{j=2}^r\xi_{-j}b_j^\dagger-
    (\sum_{j=2}^r\xi_j\xi_{-j})b_1^\dagger\right)^M|0\rangle.
    \label{drcoh2}
\end{equation}
This looks similar to the $A_{2r-1}$ and $C_r$ multinomial states, 
except that the coefficient of $b_1^\dagger$ is not independent.
The normalised $D_r$ multinomial state is
\begin{equation}
    |\hbox{\boldmath{$\eta$}};M;D_r\rangle=
    {1\over\sqrt{M!}}\left(\sum_{j=1}^r\eta_ja_j^\dagger
    +\sum_{j=1}^r\eta_{-j}b_j^\dagger\right)^M|0\rangle,
    \label{drcoh3}
\end{equation}
in which
\begin{eqnarray}
    \eta_1&=&\left(1+\sum_{j=2}^r|\xi_j|^2+\sum_{j=2}^r|\xi_{-j}|^2
    +|\sum_{j=2}^r\xi_j\xi_{-j}|^2\right)^{-{1\over2}},\
    \eta_j=\xi_j\eta_1,\quad \eta_{-j}=\xi_{-j}\eta_1,\ 2\leq j\leq 
    r,\nonumber\\
     \eta_{-1}&=&-(\sum_{j=2}^r\xi_j\xi_{-j})\eta_1,
    \label{drcohpar}
\end{eqnarray}
satisfying the condition
\begin{displaymath}
    \sum_{j=1}^r(|\eta_j|^2+|\eta_{-j}|^2)=1.
\end{displaymath}

Let us turn to the form of the probability distribution derived from the 
$D_r$ multinomial state,
which has a form similar  to that derived from the $A_r$ multinomial state. 
Similar to the $C_r$ case, $D_r$ multinomial state
predicts the multinomial distribution to the {\em number states}
with the  probabilities $|\eta_j|^2$ and $|\eta_{-j}|^2$ :
\begin{equation}
    |\langle n_1,\ldots,n_r;\bar{n}_1,\ldots,\bar{n}_r|
    \hbox{\boldmath{$\eta$}};M;D_r\rangle|^2=
    {M!\over{n_1!\cdots n_r!\bar{n}_1!\cdots 
    \bar{n}_r!}} |\eta_1|^{2n_1}\cdots|\eta_r|^{2n_r}
    |\eta_{-1}|^{2\bar{n}_1}\cdots|\eta_{-r}|^{2\bar{n}_r}.
    \label{drmulpr}
\end{equation}
By summing the
contributions from all the number states belonging to a given weight 
{\boldmath$\mu$} we obtain $D_r$ multinomial distribution:
\begin{equation}
    D_{\hbox{\boldmath{$\mu$}}}(\hbox{\boldmath{$\eta$}};M)=
    \sum_{n_j-\bar{n}_j=\mu_j}
    {M!\over{n_1!\cdots n_r!\bar{n}_1!\cdots 
        \bar{n}_r!}} |\eta_1|^{2n_1}\cdots|\eta_r|^{2n_r}
        |\eta_{-1}|^{2\bar{n}_1}\cdots|\eta_{-r}|^{2\bar{n}_r}.
    \label{drmulprform}
\end{equation}
Thus the interpretation as  ``picking up coloured balls from a pot''
is also valid.
The marked difference is that among the  probabilities
$|\eta_1|^2$, \ldots, $|\eta_r|^2$, $|\eta_{-1}|^2$, \ldots, $|\eta_{-r}|^2$,
only $2(r-1)$ of them are independent. As is clear from 
(\ref{drcohpar}), one
of the {\em dependent}   probabilities, say $|\eta_{-1}|^2$,
depends on the information of the other $\eta_{\pm j}$'s including their 
phases 
(or more precisely
$\xi_j$'s) , not $|\eta_{\pm j}|^2$'s. We believe that this is a novel feature
not encountered in any classical probability distributions.
We may say that the $D_r$ multinomial distribution has 
non-classical (or quantum) features.

\section{$B_r$ Multinomial States}
\setcounter{equation}{0}

The Dynkin diagram of $B_r$ is obtained from that of $D_{r+1}$ by 
folding the two tails.\\
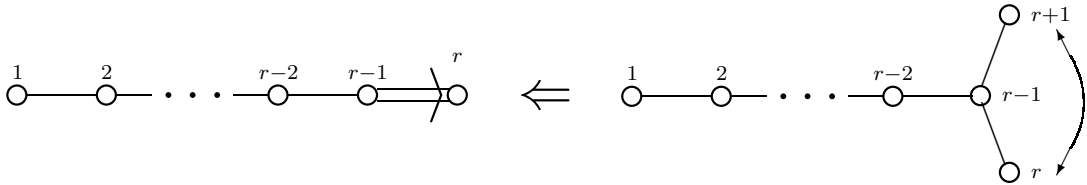
\begin{picture}(420,100)
\put(0,50){$\eddanirs1.2.{r-2}.{r-1}.r. \Leftarrow
\dddnu{1}.{2}.{r-2}.{r-1}.{r}.{r+1}.$}
\put(415,70){\vector(-1,2){5}}
\put(415,35){\vector(-1,-2){5}}
\qbezier(415,35)(425,52)(415,70)
\end{picture}\\
Thus we expect that the $B_r$ multinomial states (distributions) have
similarities with those of $D_r$ with some added new features due to 
the folding. The simple roots of $B_r$ are
\begin{equation}
    \alpha_1=e_1-e_2,\quad \alpha_2=e_2-e_3,\ \cdots,\ 
    \alpha_{r-1}=e_{r-1}-e_r,\quad \alpha_r=e_r.
    \label{brsimroot}
\end{equation}
The positive roots are
\begin{equation}
    e_j-e_k,\qquad (j<k), \qquad e_j+e_k, \qquad e_j.
    \label{brposroot}
\end{equation}
There are $2r(r-1)$ long roots and $2r$ short  roots ($\pm e_j$) and the 
dimension of $B_r$ algebra is $2r^2+r$, the same as $C_r$. 
The fundamental weights are
\begin{equation}
    \lambda_1=e_1,\qquad \lambda_2=e_1+e_2,\qquad \ldots
    \label{brwei}
\end{equation}
As before we consider the irreducible representation with the highest weight
\begin{equation}
    \mu=M\lambda_1=Me_1.
    \label{brhighwei}
\end{equation}
Let us denote this representation $\rho_B^M$ and the corresponding 
vector space by $V_B^M$.
Weyl's dimension formula gives
\begin{equation}
	\mbox{dim}(V_B^M)={M+2r-2 \choose 2r-2}\times{2M+2r-1\over{2r-1}}.
	\label{brdimfom}
\end{equation}
This representation is realised in a restricted Fock space denoted by 
$F_{2r+1}^M$:
\begin{equation}
F_{2r+1}^M;\quad |n_0,n_1,\ldots,n_r;\bar{n}_1,\ldots,\bar{n}_r\rangle,
\qquad n_0+n_1+\cdots+n_r+\bar{n}_1+\cdots+\bar{n}_r=M,
\label{f2rpfock}
\end{equation}
which is generated by $2r+1$ bosonic oscillators
\begin{displaymath}
a_0, a_1,\ldots,a_r,\quad b_1,\ldots,b_r.
\end{displaymath}
As in the $D_r$ case, by comparing the dimensions of the bosonic Fock 
space
\begin{equation}
\mbox{dim}(F_{2r+1}^M)={M+2r \choose 2r}={M+2r \choose M}
\label{f2rpdim}
\end{equation}
with the dimensions of $V_B^M$ (\ref{brdimfom}), we find
\begin{eqnarray}
\mbox{dim}(F_{2r+1}^M)&=&\mbox{dim}(V_B^M)+\mbox{dim}(F_{2r+1}^{M-2})
\nonumber\\
&=&  \mbox{dim}(V_B^M)+\mbox{dim}(V_B^{M-2})+\cdots,
\label{brdimrel}
\end{eqnarray}
which means that the bosonic Fock space $F_{2r+1}^M$ contains several
irreducible representations $\rho_B^L$ with different highest 
weights ($L=M,M-2,\ldots,$).

\bigbreak
Similarly to the $A_r$ case, the generators corresponding to various 
roots have the following 
forms:
\begin{eqnarray}
    X_{(j,-k)} & = & a_j^\dagger a_k-b_k^\dagger b_j,
    \nonumber  \\
    X_{(j,k)} & = & a_j^\dagger b_k-a_k^\dagger b_j,\quad
    X_{(-j,-k)}  =  b_j^\dagger a_k-b_k^\dagger a_j,
    \nonumber  \\
    X_{(j,0)} & = & a_j^\dagger a_0-a_0^\dagger b_j,\quad 
    X_{(-j,-j)}  =  a_0^\dagger a_j-b_j^\dagger a_0,
    \label{brgenform}
\end{eqnarray}
in which, as in the $C_r$ case, we use the notation:
\begin{eqnarray}
    X_{(j,-k)} & \Leftrightarrow & e_j-e_k,
    \nonumber  \\
    X_{(j,k)} &  \Leftrightarrow & e_j+e_k, \quad
    X_{(-j,-k)}   \Leftrightarrow  -e_j-e_k,
    \nonumber\\
    X_{(j,0)} &  \Leftrightarrow & e_j, \qquad \quad
    X_{(-j,0)}   \Leftrightarrow  -e_j.
    \label{brgennot}
\end{eqnarray}
The commutation relations are easily verified as in the previous cases.
The quadratic Casimir operator is
\begin{equation}
    C_2=N_{tot}(N_{tot}+2r-1)-4K^\dagger K,\qquad 
    N_{tot}=a_0^\dagger a_0+\sum_{j=1}^r(a_j^\dagger a_j+b_j^\dagger b_j),
    \label{brc2}
\end{equation}
in which $K$ and $K^\dagger$ are quadratic operators in the
oscillators
\begin{equation}
K={1\over2}a_0^2+\sum_{j=1}^r a_jb_j,\qquad
K^\dagger={1\over2}(a_0^\dagger)^2+\sum_{j=1}^M a_j^\dagger b_j^\dagger.  
\label{brKdefs}
\end{equation}
As in the $D_r$ cases, $K$ and $K^\dagger$ commute with all the above 
generators including those belonging to the Cartan subalgebra.
The decomposition of the restricted bosonic Fock space into the 
irreducible representation spaces goes in parallel with the $D_r$ case:
\begin{equation}
F_{2r+1}^M=V_B^M\oplus V_B^{M-2}\oplus\cdots V_B^1 (V_B^0),
\label{brrepdec}
\end{equation}
in which the vector space $V_B^M$ is obtained from the highest weight 
state
\begin{equation}
	{1\over\sqrt{M!}}(a_1^\dagger)^M|0\rangle=|0,M,0,\ldots;0,\ldots,0\rangle,
	\label{brhigh2}
\end{equation}
by applying the negative root generators successively.
The $j$-th vector space in the right hand side $V_B^{M-2(j-1)}$
is obtained from the highest weight state
\begin{equation} 
{(a_1^\dagger)^{M-2(j-1)}\over\sqrt{(M-2(j-1))!}}(K^\dagger)^{j-1}|0
\rangle,
\label{brsubsp}
\end{equation}
in a similar way.
As in the $D_r$ cases, $K$ and $K^\dagger$ annihilate all the states 
in $V_B^M$.
Thus the quadratic Casimir operator takes the value $C_2=M(M+2r-1)$ 
in the highest weight 
representation (\ref{brhighwei}),(\ref{brhigh2}).

\bigskip
One great difference between the $D_r$ and $B_r$ cases is the 
correspondence between the number states and weights.
In the $B_r$ case
\begin{equation}
    |n_0,n_1,\ldots,n_r;\bar{n}_1,\ldots,\bar{n}_r\rangle \Rightarrow
    \mu=\sum_{j=1}^r(n_j-\bar{n}_j)e_j.
    \label{brnumwei}
\end{equation}
Namely, $n_0$, the number of $a_0$ quanta, has no effects on the 
weights.

\bigskip
The $B_r$ coherent states can be constructed in a  way similar to the
$D_r$ cases. The generators having non-trivial action on the highest 
weight states are
\begin{equation}
    X_{(-1,j)},\quad  
    X_{(-1,-j)},\quad 2\leq j\leq r, \quad \mbox{and}\ \ X_{(-1,0)},
    \label{brnontri}
\end{equation}
which commute among themselves, since the sum of the corresponding roots
are no longer roots. They constitute one half of the generators 
corresponding to the quotient space
\begin{displaymath}
    SO(2r+1)/U(1)\times SO(2r-1),
\end{displaymath}
having the dimension
\begin{displaymath}
	2(2r-1).
\end{displaymath}
In terms of $2r-1$ complex parameters
\begin{equation}
   \xi_0,\quad \xi_j,\quad   \xi_{-j}, \quad 2\leq j\leq r,
    \label{brpara}
\end{equation}
we define a linear combination of the non-trivial generators 
(\ref{brnontri}) as
\begin{equation}
	T=\xi_0X_{(-1,0)}+\sum_{j=2}^r \xi_jX_{(-1,j)} 
	+\sum_{j=2}^r\xi_{-j}X_{(-1,-j)}.
	\label{brnontricom}
\end{equation}
Then the un-normalised coherent state is expressed as
\begin{equation}
	\exp[T](a_1^\dagger)^M|0\rangle,
	\label{brcoh1}
\end{equation}
which leads, after repeated use of the formal Taylor theorem 
(\ref{tay}), to
\begin{equation}
    \left(\xi_0a_0^\dagger+a_1^\dagger+\sum_{j=2}^r\xi_ja_j^\dagger+
    \sum_{j=2}^r\xi_{-j}b_j^\dagger-
    ({\xi_0^2\over2}+\sum_{j=2}^r\xi_j\xi_{-j})b_1^\dagger\right)^M|0\rangle.
    \label{brcoh2}
\end{equation}
Thus we obtain  the normalised $B_r$ multinomial state
\begin{equation}
    |\hbox{\boldmath $\eta$};M;B_r\rangle=
    {1\over\sqrt{M!}}\left(\eta_0a_0^\dagger+\sum_{j=1}^r\eta_ja_j^\dagger
    +\sum_{j=1}^r\eta_{-j}b_j^\dagger\right)^M|0\rangle,
    \label{brcoh3}
\end{equation}
in which
\begin{eqnarray}
    \eta_1&=&\left(1+\sum_{j=2}^r|\xi_j|^2+\sum_{j=2}^r|\xi_{-j}|^2
    +|{\xi_0^2\over2}+\sum_{j=2}^r\xi_j\xi_{-j}|^2\right)^{-{1\over2}},\quad
    \eta_0=\xi_0\eta_1,\nonumber\\
    \eta_j&=&\xi_j\eta_1,\ \eta_{-j}=\xi_{-j}\eta_1,\ 2\leq j\leq 
    r,\
     \eta_{-1}=-({\xi_0^2\over2}+\sum_{j=2}^r\xi_j\xi_{-j})\eta_1,
    \label{brcohpar}
\end{eqnarray}
satisfying the condition
\begin{displaymath}
   |\eta_0|^2+ \sum_{j=1}^r(|\eta_j|^2+|\eta_{-j}|^2)=1.
\end{displaymath}

\bigskip
Let us turn to the probability distribution.
The $B_r$ multinomial states give multinomial distribution to the {\em 
number states} with  probabilities
$|\eta_0|^2$, $|\eta_j|^2$ and $|\eta_{-j}|^2$ :
\begin{eqnarray}
    &&|\langle n_0,n_1,\ldots,n_r;\bar{n}_1,\ldots,\bar{n}_r|
    \hbox{\boldmath $\eta$};M;B_r\rangle|^2\label{brmulpr}\\
    &&\hspace*{1cm}=
    {M!\over{n_0!n_1!\cdots n_r!\bar{n}_1!\cdots 
    \bar{n}_r!}} |\eta_0|^{2n_0}|\eta_1|^{2n_1}\cdots|\eta_r|^{2n_r}
    |\eta_{-1}|^{2\bar{n}_1}\cdots|\eta_{-r}|^{2\bar{n}_r}.
    \nonumber
\end{eqnarray}
By summing the
contributions from all the number states belonging to a given weight 
{\boldmath $\mu$} we obtain the $B_r$ multinomial distribution:
\begin{eqnarray}
   && B_{\hbox{\boldmath $\mu$}}(\hbox{\boldmath $\eta$};M) \label{brmulprform}\\
    &=&
    \sum_{n_j-\bar{n}_j=\mu_j}
    {M!\over{n_0!n_1!\cdots n_r!\bar{n}_1!\cdots 
        \bar{n}_r!}} |\eta_0|^{2n_0}|\eta_1|^{2n_1}\cdots|\eta_r|^{2n_r}
        |\eta_{-1}|^{2\bar{n}_1}\cdots|\eta_{-r}|^{2\bar{n}_r}.
   \nonumber
\end{eqnarray}
Here let us recall that $n_0$ has no effects on the weights.
Thus the interpretation as  ``picking up coloured balls from a pot''
is also valid but with a slight modification.
In the pot we have $2r+1$ types of balls, among them $r$ different 
colours and each colour has ``positive'' and ``negative'' types.
There are also ``colourless'' (or ``dummy'') balls. They have 
 probabilities $|\eta_j|^2$, $|\eta_{-j}|^2$ 
($j=1,\ldots,r$) and $|\eta_0|^2$.
We pick up total of $M$ balls and ask the probability distribution
of the ``net'' number of coloured balls (or weights).
It is given by the $B_r$ multinomial distribution.
As in the $D_r$ multinomial distribution,  among the  probabilities
$|\eta_0|^2$,$|\eta_1|^2$, \ldots, $|\eta_r|^2$, $|\eta_{-1}|^2$, \ldots, $|\eta_{-r}|^2$,
only $2r-1$ of them are independent. As is clear from 
(\ref{brcohpar}), one
of the {\em dependent}   probabilities, say $|\eta_{-1}|^2$,
depends on the information of the other $\eta_{\pm j}$'s including their 
phases. 
The existence of the ``colourless'' balls (or dummy elements) 
and the ``quantum'' nature of $\eta_{-1}$ are novel features
of the $B_r$ multinomial distributions.


\section{Summary}
Starting from the fact established in our previous work \cite{fus7} that
the coherent states of the Heisenberg-Weyl, 
 $su(2)$, $su(r+1)$, $su(1,1)$ and $su(r,1)$ 
algebras in certain symmetric (bosonic) representations
give the well-known probability distributions, the 
Poisson, binomial, multinomial distributions with their ``negative'' 
counterparts, we have proceeded to the second stage in the study of  
``quantum probability''.
By reversing the logic, we have obtained
new probability distributions  based on the coherent states of
the classical algebras $B_r$, $C_r$ and $D_r$ in symmetric (bosonic) 
representations.
These new probability distributions have similar features as  the
multinomial distributions related with $A_r$ algebra.
They also possess several new features reflecting their Lie algebraic
and ``quantum''   backgrounds.
As byproducts, simple proofs and interpretation of some addition theorems of
Hermite polynomials are obtained
 based on the `coordinate' representation 
of the (negative) multinomial states, the coherent states of 
$su(r+1)$ ($su(r,1)$) algebra in symmetric representations.


\section*{Acknowledgements}

We thank R.\,A.\, Askey and K.\,Aomoto for useful comments and references 
of generalised Mehler formula.
We thank A.\,Bordner for reading and improving the text.
H.\,C.\,F is grateful to the Japan
Society for the Promotion of Science (JSPS) for the fellowship.
He is also supported in part by the National Science
Foundation of China.

\section*{Appendix \ \ Addition Theorems II}
\setcounter{equation}{0}
\renewcommand{\theequation}{A.\arabic{equation}} 

In this appendix we show a simple proof and interpretation of another 
type of addition theorems of Hermite polynomials.
These theorems are non-compact counterparts of the theorems presented 
in section 3.3. They are obtained from the coordinate representation of
the negative  binomial and negative multinomial states, i.e., the 
coherent states of the $su(1,1)$ and $su(r,1)$ in  symmetric 
representations.
The theorem corresponding to the negative binomial states reads
\begin{eqnarray}
	 &  & (1-\eta^2)^{-M/2}e^{x_0^2-{(x_0-\eta 
x_1)^2\over{1-\eta^2}}}H_{M-1}\left({x_0-\eta 
x_1\over\sqrt{1-\eta^2}}\right)
	\nonumber  \\
	 & = & \sum_{n=0}^\infty{(\eta/2)^n\over{n!}}H_{n+M-1}(x_0)H_n(x_1),
	\label{secadd}
\end{eqnarray}
in which $\eta$ is a complex parameter $|\eta|<1$. 
This addition theorem  is known as generalised Mehler formula
\cite{Mill,MacB}\ but is not found in the standard mathematics 
reference texts, except  for  the simplest case 
with $M=1$ which is well-known  as Mehler formula 
(see, for example, p194 of \cite{emot}). For a 
detailed  characterisation of the negative binomial (multinomial) 
distributions in terms of Lie algebras, we refer to our previous work
\cite{fus7}. 

\bigbreak
Let us begin with the negative binomial distribution (here $\eta\in 
{\mathbf R}$ for simplicity):
\begin{equation}
   B_n^-(\eta ;M )={M+n-1 \choose 
   n}\eta^{2n}(1-\eta^2)^M,\quad n=0,1,\ldots,
   \label{nbd}
\end{equation}
which describes the probability distribution of the ``waiting time''
\cite{Feller}.
Suppose we play Bernoulli's trial of success and failure in which
the  probability of {\em failure} is $0<\eta^2<1$.
The probability distribution for $n$, such that the (preset) $M$-th ($M\ge1$, 
integer) success turns out at the $M+n$-th trial, is given by the above
formula (\ref{nbd}).
We follow the examples of the previous sections and construct the
``probability amplitude'' of the negative binomial distribution.
We choose the following restricted bosonic Fock space built by two 
bosonic oscillators:
\begin{eqnarray}
     [a_j,a_k^\dagger]&=&\delta_{jk},\quad   a_j|0\rangle=0,\quad j,k=0,1, 
    \nonumber\\
     |n_0;n_1\rangle&=&{a_0^{\dagger n_0}a_1^{\dagger n_1}\over\sqrt{n_0!n_1!}}
    |0\rangle,\quad n_0-n_1=M-1,\quad n\ge0.
    \label{twobosfock2}
\end{eqnarray}
Here $n_0$ is the total number of trials except for the final one and $n_1$ is 
the number of failures (the final trial is always a success, by definition).
Obviously this Fock space is infinite dimensional.
We look for a state $|\eta;M\rangle^-$ such that
\begin{displaymath}
	|\langle n_0;n_1|\eta;M\rangle^-|^2=B_{n_1}^-(\eta;M).
\end{displaymath}
For a special choice of the phases (cf. (\ref{pnamp})) we arrive at a 
very simple result
\begin{eqnarray}
	 |\eta;M\rangle^-& = & \sum|n_0;n_1\rangle\langle n_0;n_1|\eta;M\rangle^-
	\nonumber  \\
	 & = & 
	 (1-\eta^2)^{M\over2}\sum|n_0;n_1\rangle\eta^n\sqrt{n_0!\over{n_1!(M-1)!}}
	\nonumber  \\
	 & = &  (1-\eta^2)^{M\over2}\sum_{n_1=0}^\infty{(\eta a_0^\dagger 
	 a_1^\dagger)^{n_1}\over{n_1!}}
	 {(a_0^\dagger)^{M-1}\over\sqrt{(M-1)!}}|0\rangle
	\nonumber  \\
	 & = &  (1-\eta^2)^{M\over2}e^{\eta a_0^\dagger 
	 a_1^\dagger}|M-1;0\rangle.
	\label{negbst1}
\end{eqnarray}
This  is called the negative binomial state \cite{fus6,fus7}.
This is exactly an $su(1,1)$ coherent state as we will see presently.
The $su(1,1)$ algebra is realised in the above Fock space as
\begin{eqnarray}
    K_+&=&a_0^\dagger a_1^\dagger,\quad K_-=a_0a_1,
    \quad K_0={1\over2}(N_0+N_1+1),\quad N_j=a_j^\dagger a_j,
    \nonumber\\
   \ [K_+,K_-]&=&-2K_0,\quad [K_0,K_\pm]=\pm K_\pm.
    \label{11alge}
\end{eqnarray}
The lowest weight state is $|M-1;0\rangle$:
\begin{equation}
	K_-|M-1;0\rangle=0,\quad K_0|M-1;0\rangle={M\over2}|M-1;0\rangle,
	\label{su11low}
\end{equation}
which gives rise to the discrete irreducible representation with 
Bargman index $M/2$. Thus the un-normalised coherent state is 
($\eta\in{\mathbf C}$)
\begin{equation}
			e^{\eta K_+}|M-1;0\rangle=e^{\eta a_0^\dagger 
			a_1^\dagger}|M-1;0\rangle,
			\label{negbst2}
\end{equation}
which has the same form as given in (\ref{negbst1}).

\bigbreak
Next we take the coordinate representation of the above negative
binomial state:
\begin{displaymath}
	\langle x_0;x_1|e^{\eta a_0^\dagger a_1^\dagger}|M-1;0\rangle
\end{displaymath}
and evaluate it in two different ways. The first is to simply expand
the exponential and use the formula (\ref{xrep}):
\begin{equation}
	\langle x_0;x_1|e^{\eta a_0^\dagger a_1^\dagger}|M-1;0\rangle
	 ={e^{-{1\over2}(x_0^2+x_1^2)}\over{\pi^{1/2}\sqrt{(M-1)!}}}
	 \sum_{n=0}^\infty{(\eta/2)^n\over{n!}}
	 H_{n+M-1}(x_0)H_n(x_1),
	\label{xrepnbs1}
\end{equation}	
which corresponds to the right hand side of (\ref{secadd}).

The second is to use the coordinate representation of the creation 
operators
\begin{displaymath}
	a_j^\dagger={1\over\sqrt2}(x_j-{\partial\over{\partial x_j}})
	=-{1\over\sqrt2}e^{{1\over2}x_j^2}D_je^{-{1\over2}x_j^2},\quad 
	D_j={\partial\over{\partial x_j}},\quad j=0,1,
\end{displaymath}
to obtain
\begin{displaymath}
	\langle x_0;x_1|e^{\eta a_0^\dagger a_1^\dagger}|M-1;0\rangle
		 ={(-1)^{M-1}\over{\pi^{1/2}\sqrt{(M-1)!}}}\,e^{{1\over2}(x_0^2+x_1^2)}
		 e^{\eta D_0 D_1/2}D_0^{M-1}e^{-(x_0^2+x_1^2)}.
\end{displaymath}
By applying the formal Taylor theorem (\ref{tay}) with respect to $x_1$
by treating $\eta D_0$ as a parameter, we obtain
\begin{eqnarray}
	 &  & 	\langle x_0;x_1|e^{\eta a_0^\dagger a_1^\dagger}|M-1;0\rangle
	\nonumber  \\
	 & = & {(-1)^{M-1}e^{{1\over2}(x_0^2+x_1^2)}\over{\pi^{1/2}\sqrt{(M-1)!}}}
	 \,D_0^{M-1}e^{-(x_1+\eta D_0/2)^2}e^{-x_0^2}
	\nonumber  \\
	 & = & {(-1)^{M-1}e^{{1\over2}(x_0^2-x_1^2)}\over{\pi^{1/2}\sqrt{(M-1)!}}}
	 {1\over\sqrt{1-\eta^2}}\,e^{-\eta x_1 
	 D_0}D_0^{M-1}e^{-{x_0^2\over{1-\eta^2}}},
	\label{xrepnbs2}
\end{eqnarray}
which gives a scaled ($1/\sqrt{1-\eta^2}$) and shifted ($-\eta x_1$) 
Hermite polynomial ($H_{M-1}$) by Rodrigues formula (\ref{rodr}):
\begin{equation}
	\mbox{r.h.s of (\ref{xrepnbs2})}=
	{1\over{\pi^{1/2}\sqrt{(M-1)!}}}(1-\eta^2)^{-{M\over2}}
	e^{{1\over2}x_0^2}e^{-{(x_0-\eta x_1)^2\over{1-\eta^2}}}
	H_{M-1}\left({x_0-\eta x_1\over\sqrt{1-\eta^2}}\right).
	\label{xrepnbs3}
\end{equation}
Here use is made of a simple formula
\begin{displaymath}
	e^{tD_0^2}e^{-x_0^2}={1\over\sqrt{1+4t}} e^{-{x_0^2\over{1+4t}}},\quad
	|t|<{1\over2},
\end{displaymath}
which can be proved, for example,  by taking the Fourier transform.
By comparing (\ref{xrepnbs2}) and (\ref{xrepnbs3}) we arrive at the 
addition theorem of Hermite polynomials given above (\ref{secadd}).
It should be remarked that the generalised Mehler formula (\ref{secadd})
is also obtained from Mehler formula ($M=1$) by differentiating with 
respect to $x_0$ $M-1$ times.

\bigbreak
Generalisation to the negative multinomial distribution
\begin{eqnarray}
    M_{\mathbf{n}}^-(\hbox{\boldmath $\eta$}\,;M)&=&
    (1-\hbox{\boldmath $\eta$}^2)^M
    {(M+n_1+\cdots+n_r-1)!\over{\mathbf 
    {n}!(M-1)!}}\eta_1^{2n_1}\cdots\eta_r^{2n_r},
    \label{negmulnomdis}\\
    \mathbf{n}&=&(n_0,n_1,\ldots,n_r),\quad 
     \quad \hbox{\boldmath $\eta$}=(\eta_1,\ldots,\eta_r)\in\mathbf 
    {R}^r,\label{negmulnot}\\
     0&<&\hbox{\boldmath $\eta$}^2=\eta_1^2+\cdots+\eta_r^2<1,
    \nonumber
\end{eqnarray}
is rather straightforward. We introduce a restricted Fock space
generated by $r+1$ oscillators:
\begin{eqnarray}
    [a_j,a_k^\dagger]&=&\delta_{jk},\quad a_j| 0\rangle=0,
    \quad j=0,1,\ldots,r,\label{negmulosci}\\ 
     |n_0;n_1,\ldots,n_r\rangle&=&{(a_0^\dagger)^{n_0}(a_1^\dagger)^{n_1}
     \cdots(a_r^\dagger)^{n_r}\over\sqrt{n_0!n_1!\cdots n_r!}}|0\rangle,
     \quad n_0-(n_1+\cdots+n_r)=M-1.
    \nonumber
\end{eqnarray}
Then the ``square root'' of the negative multinomial distribution is
\begin{equation}
	 |\hbox{\boldmath $\eta$}\,;M\rangle^-=(1-\hbox{\boldmath 
	 $\eta$}^2)^{M\over2} e^{a_0^\dagger(\sum_{j=1}^r\eta_j a_j^\dagger)}
	 |M-1;0,\ldots,0\rangle,
 	\label{negmulst1}
\end{equation}
which is an $su(r,1)$ coherent state in an irreducible symmetric 
 representation with the lowest weight state
\begin{equation}
	 |M-1;0,\ldots,0\rangle.
 	\label{negloweist}
\end{equation}
The generators are
 \begin{eqnarray}
K_{+j}&=&a_0^\dagger a_j^\dagger,\quad K_{-k}=a_0 a_k,\quad 1\leq 
j,k\leq r,\nonumber\\
K_{jk}&=&a_j^\dagger a_k\quad (j\neq k\neq0),\quad N_j=a_j^\dagger a_j.
\label{sur1lin}
\end{eqnarray}
It is easy to see that they leave the combination
\begin{displaymath}
	\Delta\equiv N_0-(N_1+\cdots+N_r)
\end{displaymath}
and the above Fock space (\ref{negmulosci}) invariant.
Among the above generators the following  $r$ generators have 
non-trivial action on the lowest weight state (\ref{negloweist})
\begin{equation}
	K_{+j}=a_0^\dagger a_j^\dagger,\quad j=1,\ldots,r.
	\label{negnontri}
\end{equation}
Thus in terms of $r$ complex parameters $\eta_1$,\ldots,$\eta_r$, 
satisfying the condition
\begin{equation}
	|\hbox{\boldmath $\eta$}|^2=\sum_{j=1}^r|\eta_j|^2<1,
	\label{negmulpara}
\end{equation}
we obtain an un-normalised negative multinomial state
\begin{equation}
	e^{\sum_{j=1}^r\eta_j K_{+j}}|M-1;0,\ldots,0\rangle=
	e^{a_0^\dagger(\sum_{j=1}^r\eta_j a_j^\dagger)} 
	|M-1;0,\ldots,0\rangle,
	\label{negmulst2}
\end{equation}
which has the same form as (\ref{negmulst1}).
By evaluating the coordinate representation of the above state 
(\ref{negmulst2}) in two different ways, we obtain another form of
addition theorem of Hermite polynomials:
\begin{eqnarray}
	 &  & (1-\hbox{\boldmath $\eta$}^2)^{-M/2}
	 e^{x_0^2-{(x_0-\eta_1 x_1-\cdots-\eta_rx_r)^2
	 \over{1-\eta_1^2\cdots-\eta_r^2}}}
	 H_{M-1}\left({x_0-\eta_1 x_1-\cdots-\eta_rx_r\over
	 \sqrt{1-\eta_1^2\cdots-\eta_r^2}}\right)
	\nonumber  \\
	 & = & \sum_{n_j=0}^\infty{(\eta_1/2)^{n_1}\over{n_1!}}\cdots
	 {(\eta_r/2)^{n_r}\over{n_r!}}H_{M+n_1\cdots+n_r-1}(x_0)
	 H_{n_1}(x_1)\cdots
	 H_{n_r}(x_r),
	\label{thiradd}
\end{eqnarray}
One can obtain this addition theorem by combining the addition theorems
from the multinomial state (\ref{addthe1}) and that of the negative binomial
state (\ref{secadd}), which reflects the fact that the
negative multinomial state is also obtained by combining the negative
binomial state and the multinomial state.

 \bigbreak
 Before closing Appendix, let us mention another interesting form of
 addition theorems of Hermite polynomials which is obtained as a special case of
 (\ref{secadd}). By setting $x_0\equiv x$ and $x_1\equiv0$, we obtain
\begin{equation}
	 (1-\eta^2)^{-M/2}e^{-{\eta^2\over{1-\eta^2}}x^2}
	 H_{M-1}({x\over\sqrt{1-\eta^2}})
	  = \sum_{n=0}^\infty{(-\eta^2/4)^n\over{n!}}H_{2n+M-1}(x).
	\label{fouradd}
\end{equation}
Here use is made of the relations
\begin{displaymath}
	H_{2n}(0)=(-1)^n(2n-1)!!=(-1)^n1\cdot3\cdots(2n-1),\quad H_{2n+1}(0)=0.
\end{displaymath}
This form of addition theorems can also be obtained from another type 
of ``coherent states'' of $su(1,1)$ algebra.
Let us take the single boson Fock space (\ref{aadag})--(\ref{vac}) with the 
basis 
\begin{math}
	\{|n\rangle,\ n=0,1,\ldots,\}
\end{math}
generated by $a$ and $a^\dagger$. The $su(1,1)$ algebra is realised by
\begin{equation}
	K_+={1\over2}(a^\dagger)^2,\quad K_-={1\over2}a^2,\quad 
	K_0={1\over2}a^\dagger a+{1\over4}.
	\label{secsu11}
\end{equation}
As before evaluate an un-normalised ``coherent state''
\begin{equation}
	e^{tK_+}|M-1\rangle=e^{{t\over2}(a^\dagger)^2}|M-1\rangle,\quad |t|<1,
	\label{secsu11cohe}
\end{equation}
in two different ways ($t=-\eta^2$). The above state is known as the 
`squeezed number state' in quantum optics \cite{squnum}, for the `base state'
$|M-1\rangle$ is not of lowest weight.


\end{document}